%% file: EarlyAgeUHPC.tex
\documentclass[12pt]{article}

\input{definitions.tex}

\begin{document}


\include{segim}

\begin{center}
{\large {\bf Analysis of the Behavior of Ultra High Performance Concrete at Early Age}}
\\[9mm]
{\bf  {\bf  By \\
Lin Wan \footnote{PhD, Researcher, Department of Civil and Environmental Engineering, Northwestern University, 2145 Sheridan Rd. A120, Evanston IL, 60208 USA. E-mail: \mbox{lin.wan@u.northwestern.edu}},
Roman Wendner$^*$ \footnote{$^*$Corresponding Author: Director Christian Doppler Laboratory LiCRoFast, Department of Civil Engineering and Natural Hazards, University of Natural Resources and Life Sciences (BOKU) Vienna, Austria. E-mail: \mbox{roman.wendner@boku.ac.at}, Tel: +43 1 47654 5252},
Benliang Liang \footnote{Associate Professor, Department of Architecture and Civil Engineering, Shanghai Normal University, 100 Haisi Rd. Fengxian District, Shanghai, P.R.C. E-mail: \mbox{lbl@shnu.edu.cn}},
Gianluca Cusatis \footnote{Associate Professor, Department of Civil and Environmental Engineering, Northwestern University, 2145 Sheridan Rd. A123, Evanston IL, 60208 USA. E-mail: \mbox{g-cusatis@northwestern.edu}} 
}  }

\vspace{20mm}

A Paper Published in

Cement \& Concrete Composites 

doi: 10.1016/j.cemconcomp.2016.08.005

September, 2016

Corresponding Author

Roman Wendner

Director - Christian Doppler Laboratory LiCRoFast

Deputy Director - Institute of Structural Engineering

Department of Civil Engineering and Natural Hazards

University of Natural Resources and Life Sciences (BOKU)

Peter Jordan Strasse 82, A-1190

Vienna, Austria

Tel: +43 1 47654 5252

Fax: +43 1 47654 5299

Email: roman.wendner@boku.ac.at

Website: http://www.baunat.boku.ac.at/cd-labor

%
%
%
%
%
%
%
%
%

\end{center}

\newpage

\no {\bf   Abstract:}\\ {\sf

Ultra high performance concretes (UHPCs) are cementitious composite materials with high level of performance characterized by high compressive strength, high tensile strength and superior durability, reached by low water-to-binder ratio, optimized aggregate size distribution, thermal activation, and fiber reinforcement. In the past couple of decades, more and more UHPCs have been developed and found their ways into practice. Thus, the demand for computational models capable of describing and predicting relevant aging phenomena to assist design and planning is increasing. This paper presents the early age experimental characterization as well as the results of subsequent simulations of a typical UHPC matrix. Performed and simulated tests include unconfined compression, splitting (Brazilian), and three-point-bending tests. The computational framework is formulated by coupling a hygro-thermo-chemical (HTC) theory and a comprehensive mesoscale discrete model with formulated aging functions. The HTC component allows taking into account various types of curing conditions with varying temperature and relative humidity and predicting the level of concrete aging. The mechanical component, the Lattice Discrete Particle Model (LDPM), permits the simulation of the failure behavior of concrete at the length scale of major heterogeneities. The aging functions relate the mesoscale LDPM mechanical properties in terms of aging degree, defined in this work as the ratio between the quasi-static elastic modulus at a certain age and its asymptotic value. The obtained results provide insights in both UHPC early age mechanisms and a computational model for the analysis of aging UHPC structures. 

\section{Introduction}\no

Ultra high performance concretes (UHPCs) are cementitious composites characterized by high compressive strength, typically greater than 120 MPa (17~ksi), low water-binder ratio, optimized gradation curve, use of thermal activation, fiber reinforcement and superplasticizers. Moreover, UHPC has a discontinuous pore structure that reduces liquid ingress and permeability \cite{Gray2011}, leading to significantly enhanced durability, longer service life and lower costs for maintenance. UHPC became commercially available in the beginning of the 21st century and has been utilized in the construction industry especially for bridge applications and tall buildings around the world across North America, Europe, and Asia. 

Among various types of UHPCs developed by researchers around the globe, Ductal is the most commonly used and is commercially available in the United States, Europe and Asia. Other types of UHPCs include but are not limited to compact reinforced composite (CRC) developed by Aalborg Portland in 1986, the UHPC mix of Teichmann and Schmidt in Germany, and CorTuf developed by the U.S. Army Corps of Engineers Research and Development Center (ERDC), which all utilize water/cement ratios within the range of 0.15$\sim$0.21, fine aggregate gradation, silica fume, superplasticizers and steel fibers \cite{FHWA2013}. The inclusion of silica fume allows a pozzolanic reaction to produce additional calcium-silicate-hydrate (C-S-H), which can fill the would-have-been voids and provide a much denser matrix. The perfectly graded fine aggregate plays a similar role but as filler in minimizing the number and volume of possible voids. 

While more and more UHPCs are developed and utilized in the construction industry, what is lacking in the available literature is a model for the evolution of material properties of UHPCs at early age. This is crucial in terms of structural design, project planning, and building optimization. The construction of bridges and skyscrapers, for example, structures in which UHPC materials are utilized the most, are typically characterized by a complex sequence of construction stages which may last for months or even years. Being able to model how the material ages and at what pace each mechanical property grows could greatly help optimize the construction sequence and hereby reduce the total construction time and budget. The behavior at early age is also of relevance for the long-term performance and durability due to e.g. restraint forces \cite{Wendner14}, or cracking at early age \cite{Hubler15,Wendner15}. Consequently, such a computational framework is quintessential to the accurate performance assessment of degrading concrete structures since all long-term analyses are anchored to the construction stages \cite{Strauss13}.

Moreover, the demand for computational models that can be utilized in design is growing with the increased adoption of UHPC materials in practice. For example, Chen and Graybeal \cite{Graybeal2010} investigated the behavior of existing UHPC structural components including prestressed girders by finite element methods and showed good abilities of replicating the structural response. However, conventional constitutive formulations are unable to capture the evolution of material properties and, correspondingly, the time-dependent structural response of concrete as an aging material. This is an important aspect to consider given that the current construction practice requires shorter and shorter construction times associated with, for example, early load applications and prestressing. While the contributions of fibers are quite essential for the success of UHPCs, this paper focuses solely on a computational framework for the evolution of the matrix properties. The presence of fiber reinforcement is not directly investigated here. However, related research efforts have shown that the behavior of fiber reinforced UHPCs can be predicted based on a calibrated constitutive model for the plain concrete and an appropriate fiber concrete interaction model for the discretely placed fibers \cite{Schauffert2012I}. 

It is well known that the strength of cement based composites such as concrete increases rapidly at early age. However, the chemical and physical mechanisms behind this phenomenon are complex and consist of multiple coupled components. The cross-effects between hydration reaction, temperature evolution, and member deformation involve complex chemo-physical mechanisms that operate over a broad range of length and time scales, from nanometer to meter, and from fractions of seconds to years \cite{Ulm1998}. Notably, evolution laws for maturing concrete based on Arrhenius-type time acceleration concepts are widely supported by a good agreement with experimental data \cite{Byfors1980, Regourd1980, Ulm1995}. 

Ulm and Coussy \cite{Ulm1995} studied the thermo-chemo-mechanical coupling of concrete at early age with a formulation based upon thermodynamics of open porous media composed of a skeleton and several fluid phases saturating the porous space. It accounts explicitly for the hydration of cement by considering the thermodynamic imbalance between the chemical constituents in the constitutive modeling at the macrolevel of the material description. However, the effects from stress and temperature evolutions were neglected. Afterwards they extended the thermo-chemo-mechanical cross effects characterizing the autogenous shrinkage, hydration heat and strength growth, within the framework of chemoplasticity \cite{Ulm1998}. Cervera et. al. \cite{Cervera1999} applied the reactive porous media theory and introduced a novel aging model which accounts for the effect of curing temperature evolution featuring the aging degree as an internal variable. They suggested that the evolution of the compressive and tensile strengths and elastic moduli can be predicted in terms of the evolution of the aging degree \cite{Cervera2000, Cervera2000II}. The model considers the short-term mechanical behavior based on the continuum damage mechanics theory and the long-term mechanical behavior based upon the microprestress-solidification theory \cite{Cervera1999II}. Bernard, Ulm and Lemarchand \cite{Bernard2003} developed a multi scale micromechanics-hydration model to predict the aging elasticity of cement-based materials starting at the nano level of the C-S-H matrix. Lackner and Mang \cite{Lackner2004} proposed a 3-D material model for the simulation of early-age cracking of concrete based on the Rankine criterion formulated in the framework of multi surface chemoplasticity. Gawin, Pesavento, and Schrefler \cite{Gawin2006, Gawin2006II} proposed a solidification-type early-age model and extended it to account for coupled hygro-thermo-chemo-mechanical phenomena, which was applied to solving practical problems, such as analysis, repair, and rehabilitation of concrete structures \cite{Sciume2013, Cervenka2014}.

Di Luzio and Cusatis \cite{DiLuzio2009I,DiLuzio2009II} formulated, calibrated, and validated a hygro-thermo-chemical (HTC) model suitable for the analysis of moisture transport and heat transfer for standard as well as high performance concrete. In this study, classical macroscopic mass and energy conservation laws were formulated with humidity and temperature as primary variables and by taking into account explicitly various chemical reactions including cement hydration, silica fume reaction, and silicate polymerization \cite{DiLuzio2009I}. Furthermore, Di Luzio and Cusatis \cite{DiLuzio2013}, amalgamated the microplane model and the microprestress-solidification theory. This unified model takes into account all the most significant aspects of concrete behavior, such as creep, shrinkage, thermal deformation, and cracking starting from the initial stages of curing up to several years of age. 

While continuum mechanics and finite element solvers are broadly utilized for mechanical analysis of concrete structures at the macroscopic levels, the Lattice Discrete Particle Model (LDPM) \cite{Cusatis2011a, Cusatis2011b} provides additional insights into failure behavior of concrete at smaller length scales. LDPM simulates concrete at the length scale of coarse aggregate pieces (mesoscale) and is formulated within the framework of discrete models, which enable capturing the salient aspects of material heterogeneity while keeping the computational cost manageable \cite{Cusatis2011a}. 

The HTC model and LDPM are selected as basis for the early age mechanical model formulated in this study. On the one hand, the HTC model can comprehensively capture the hygral and thermal evolutions and chemical reactions during aging. On the other hand, LDPM can provide insights into the concrete behavior at the mesoscale level and also simulate well the mechanical behavior of concrete structures under various loads at the macroscopic level. The two models are connected by a set of proposed aging functions that form the core of the proposed HTC-LDPM coupled early age framework, allowing prediction of the behavior of concrete members at early age and beyond. In support of the calibration and validation tasks, a comprehensive experimental investigation for a UHPC at early age was also carried out. The performed tests include unconfined cylinder \& cube compression, cylinder splitting, notched and unnotched beam three-point-bending tests, and internal relative humidity measurements on cylindrical specimens. The evolution of unconfined compressive strength, tensile splitting strength, tensile flexural strength, and the associated evolving stress-strain relationships are then extracted from the experiments and presented in terms of concrete aging.

\section{Computational Framework for the Simulation of UHPC at Early Age}

The proposed hygro-thermo-chemo-mechanical early-age model for cement based concrete consists of three major components: the hygro-thermo-chemical (HTC) model, the Lattice Discrete Particle Model (LDPM), and the formulated aging functions bridging the two.

\subsection{Hygro-Thermo-Chemical (HTC) model}

The behavior of concrete at early age heavily depends on moisture content and temperature. The overall moisture transport can be described through Fick's law that expresses the flux of water mass per unit time $\mathbf{J}$ as a function of the spatial gradient of the relative humidity $h$. Moisture mass balance requires the variation in time of the water mass per unit volume of concrete to be equal to the divergence of moisture flux, $\mathrm{div}\,\mathbf{J}$. The water content $w$ is the sum of evaporable water $w_e$ (capillary water, water vapor, and absorbed water) and non-evaporable (chemically bound) water $w_n$ \cite{Powers1946, Mills1966,Panta1995}. Assuming that $w_e$ is a function of relative humidity $h$, degree of hydration $\alpha_c$, and degree of silica fume reaction $\alpha_s$, one can write $w_e = w_e (h, \alpha_c, \alpha_s)$, which represents an age-dependent sorption/desorption isotherm. Consequently, the moisture mass balance equation reads \cite{DiLuzio2009I}:

\begin{equation}
\nabla \cdot (D_h \nabla h) - \frac{\partial w_e}{\partial h} \frac{\partial h}{\partial t} - \left(\frac{\partial w_e}{\partial \alpha_c} \dot{\alpha_c} + \frac{\partial w_e}{\partial \alpha_s} \dot{\alpha_s} + \dot{w_n}\right) = 0
\end{equation}
where $D_h$ is moisture permeability. The enthalpy balance is also influenced by the chemical reactions occurring at the early age. One can write, at least for temperatures not exceeding 100~$^\circ$C \cite{Bazant1996}, 

\begin{equation}
\nabla \cdot (\lambda_t \nabla T) - \rho c_t \frac{\partial T}{\partial t} + \dot{\alpha}_cc\tilde{Q}^{\infty}_c + \dot{\alpha}_ss\tilde{Q}^{\infty}_s = 0
\end{equation}

where $\tilde{Q}^{\infty}_c$ = hydration enthalpy, $\tilde{Q}^{\infty}_s$ = latent heat of silica-fume reaction per unit mass of reacted silica-fume, $\rho$ is the mass density of concrete, $\lambda_t$ is the heat conductivity, and $c_t$ is the isobaric heat capacity of concrete. 

According to the thermodynamics based model proposed by Ulm and Coussy \cite{Ulm1995,Mazloom2004} and later revised by Cervera et al. \cite{Cervera1999}, the hydration kinetics can be described by postulating the existence of a Gibb's free energy dependent on the external temperature $T$ and the hydration extent $\chi_c$. For convenience the hydration degree can be defined as
$\alpha_c = \chi_c/\overline{\chi}^{\infty}_c$, where $\overline{\chi}^{\infty}_c$ is the theoretical asymptotic value of the hydration extent in ideal hygrometric conditions. Analogously, the asymptotic degree of hydration, can be expressed as $\alpha^{\infty}_c = \chi^{\infty}_c/\overline{\chi}^{\infty}_c$.

With the two assumptions that (1) the thermodynamic force conjugate to the hydration extent, named the \textit{chemical affinity}, is governed by an Arrhenius-type equation and (2) the viscosity governing the diffusion of water through the layer of cement hydrates is an exponential function of the hydration extent \cite{Ulm1995}, Cervera et al. \cite{Cervera1999} proposed the evolution equation for the hydration degree: $\dot{\alpha}_c = A_c(\alpha_c)e^{-E_{ac}/RT}$, and \mbox{$A_c(\alpha_c) = A_{c1}(\frac{A_{c2}}{\alpha^{\infty}_c} + \alpha_c)(\alpha^{\infty}_c - \alpha_c)e^{-\eta_c\alpha_c/\alpha^{\infty}_c}$}, where $A_c(\alpha_c)$ is the \textit{normalized chemical affinity}, $E_{ac}$ is the hydration activation energy, $R$ is the universal gas constant, and $\eta_c$, $A_{c1}$, and $A_{c2}$ are material parameters. To account for the situation that the hydration process slows down and may even stop if the relative humidity decreases below a certain value, the equation can be rewritten as: $\dot{\alpha}_c = A_c(\alpha_c)\beta_h(h)e^{-E_{ac}/RT}$, where $\beta_h(h) = [1 + (a - ah)^b]^{-1}$. The function $\beta_h(h)$ is an empirical function that was first proposed for the definition of the equivalent hydration period by Ba\v{z}ant and Prasannan \cite{Bazant1989a}. The parameters $a$ and $b$ can be calibrated by experimental data but values $a$ = 5.5 and $b$ = 4 can be generally adopted \cite{Bazant1989a, Gawin2006}.

The theory adopted for cement hydration can be as well utilized for silica fume (SF) reaction since the kinetics of a pozzolanic reaction can also be assumed to be diffusion controlled. Accordingly, the degree of SF reaction, $\alpha_s$, is introduced, $\dot{\alpha}_s = A_s(\alpha_s)e^{-E_{as}/RT}$, where $A_s(\alpha_s) = A_{s1}(\frac{A_{s2}}{\alpha^{\infty}_s} + \alpha_s)(\alpha^{\infty}_s - \alpha_s)e^{-\eta_s\alpha_s/\alpha^{\infty}_s}$, is the SF normalized affinity, $E_{as}$ is the activation energy of SF reaction, and $\alpha^{\infty}_s$ is the asymptotic value of the SF reaction degree \cite{DiLuzio2009I}. $E_{as}/R = 9700$~K can be generally assumed \cite{Bentz1998}. The rate of SF reaction degree is assumed not to depend on $h$ since the pozzolanic reaction continues even at very low relative humidities \cite{Jensen1996}. 

Hydration degree and silica fume reaction degree can also be combined to define a total reaction degree \cite{DiLuzio2013} as
\begin{equation}
\alpha(t)=\frac{\alpha_c(t)c\tilde{Q}_c^\infty + \alpha_s(t)s\tilde{Q}_s^\infty}{\alpha_c^\infty c\tilde{Q}_c^\infty + \alpha_s^\infty s\tilde{Q}_s^\infty}
\end{equation} 
A detailed discussion of the calibration procedure for the HTC model can be found in Di Luzio and Cusatis \cite{DiLuzio2009II}.

\subsection{Lattice Discrete Particle Model}

In 2011, Cusatis and coworkers \cite{Cusatis2011a,Cusatis2011b} developed the Lattice Discrete Particle Model (LDPM), a mesoscale discrete model that simulates the mechanical interaction of coarse aggregate pieces embedded in a cementitious matrix (mortar). The geometrical representation of the concrete mesostructure is constructed through the following steps. First, the coarse aggregate pieces, assumed to have spherical shapes, are introduced into the concrete volume by a try-and-reject random procedure. Secondly, nodes as zero-radius aggregate pieces are randomly distributed over the external surfaces to facilitate the application of boundary conditions. Thirdly, a three-dimensional domain tessellation, based on the Delaunay tetrahedralization of the generated particle centers, creates a system of polyhedral cells (see Fig.~\ref{ldpm}) interacting through triangular facets and a lattice system composed by the line segments connecting the aggregate centers. The full description of LDPM geometry is reported in Cusatis et al. \cite{Cusatis2011a,Cusatis2011b}.

\begin{figure}[h]
  \centering 
  \includegraphics[width=2.5 in]{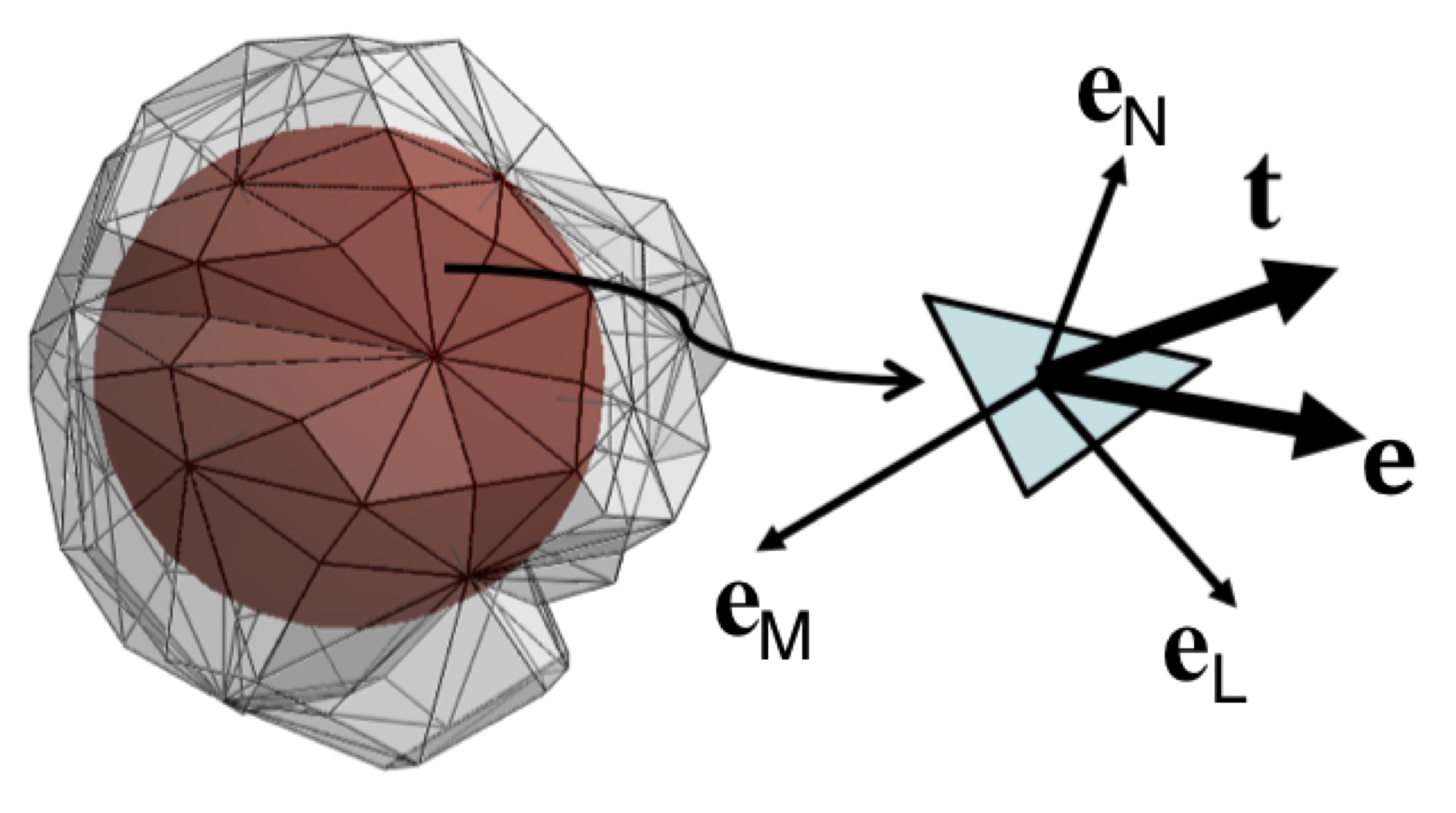}
\caption{One LDPM Cell around an aggregate piece. }
    \label{ldpm}
\end{figure}

In LDPM, rigid body kinematics are used to describe the deformation of the lattice particle system and the displacement jump, $\llbracket \mathbf{u}_{C} \rrbracket$, at the centroid of each facet is used to define measures of strain as 
\begin{equation}
\label{eps} e_{N}=\frac{\mathbf{n}^\mathrm{T} \llbracket \mathbf{u}_{C} \rrbracket}{\ell};
\;\;\; e_{L}=\frac{\mathbf{l}^\mathrm{T} \llbracket \mathbf{u}_{C} \rrbracket} {\ell};
\;\;\; e_{M}=\frac{ \mathbf{m}^\mathrm{T} \llbracket \mathbf{u}_{C} \rrbracket}{\ell}
\end{equation}
where $\ell=$ interparticle distance; and $\mathbf{n}$, $\mathbf{l}$, and
$\mathbf{m}$, are unit vectors defining a local system of reference attached to each facet. It was recently demonstrated that the strain definitions in Eq. \ref{eps} correspond to the projection into the local system of references of the strain tensor typical of continuum mechanics \cite{DCM,HOMO,HIGH_MP}.

Next, a vectorial constitutive law governing the behavior of the material is imposed at the centroid of each facet. In the elastic regime, the normal and shear stresses are proportional to the corresponding strains: $t_{N}= E_N e^*_{N} =E_N (e_{N}-e^0_{N});~ t_{M}= E_T e^*_{M} = E_T (e_{M}-e^0_{M});~ t_{L}= E_T e^*_{L} = E_T (e_{L}-e^0_{L})$, where $E_N=E_0$, $E_T=\alpha E_0$, $E_0=$ effective normal modulus, and $\alpha=$ shear-normal coupling parameter; and $e^0_{N}$, $e^0_{M}$, $e^0_{L}$ are mesoscale eigenstrains that might arise from a variety of phenomena such as, but not limited to, thermal expansion, creep, shrinkage, and deterioration mechanisms (e.g. alkali-silica reaction).

 For stresses and strains beyond the elastic limit, the LDPM formulation considers the following nonlinear mesoscale phenomena \cite{CuBaCe2003a, CuBaCe2003b, Cusatis2011a}: (1) fracture and cohesion; (2) compaction and pore collapse; and (3) friction.

\textbf{Fracture and cohesion due to tension and tension-shear.} For tensile loading ($e^*_N>0$), the fracturing behavior is formulated through an effective strain, $e^* = \sqrt{e_N^{*2}+\alpha (e_M^{*2} + e_L^{*2})}$, and stress, $t = \sqrt{{ t _{N}^2+  (t _{M}^2+t _{L}^2) / \alpha}}$, which define the normal and shear stresses as \mbox{$t _{N}= e_N^*(t / e^*)$}; \mbox{$t _{M}=\alpha e^*_{M}(t / e^*)$}; \mbox{$t _{L}=\alpha e^*_{L}(t / e^*)$}. The effective stress $t$ is incrementally elastic ($\dot{t}=E_0\dot{e}$) and must satisfy the inequality $0\leq t \leq \sigma _{bt} (e, \omega) $ where $\sigma_{bt} = \sigma_0(\omega) \exp \left[-H_0(\omega)  \langle e-e_0(\omega) \rangle / \sigma_0(\omega)\right]$, $\langle x \rangle=\max \{x,0\}$, and $\tan(\omega) =e^* _N / \sqrt{\alpha} e^* _{T}$ = $t_N \sqrt{\alpha} / t_{T}$, and $e_T^*=\sqrt{e_M^{*2} + e_L^{*2}}$. The post peak softening modulus is defined as $H_{0}(\omega)=H_{t}(2\omega/\pi)^{n_{t}}$, where $H_{t}$ is the softening modulus in pure tension ($\omega=\pi/2$) expressed as $H_{t}=2E_0/\left(\ell_t/\ell-1\right)$; $\ell_t=2E_0G_t/\sigma_t^2$; $\ell$ is the length of the tetrahedron edge; and $G_t$ is the mesoscale fracture energy. LDPM provides a smooth transition between pure tension and pure shear ($\omega=0$) with parabolic variation for strength given by $\sigma_{0}(\omega )=\sigma _{t}r_{st}^2\Big(-\sin(\omega) + \sqrt{\sin^2(\omega)+4 \alpha \cos^2(\omega) / r_{st}^2}\Big) / [2 \alpha \cos^2(\omega)]$, where $r_{st} = \sigma_s/\sigma_t$ is the ratio of shear strength to tensile strength. 

\textbf{Compaction and pore collapse from compression.} Normal stresses for compressive loading ($e^*_N<0$) must satisfy the inequality $-\sigma_{bc}(e_D, e_V)\leq t_N \leq 0$, where $\sigma_{bc}$ is a strain-dependent boundary depending on the volumetric strain, $e_V$, and the deviatoric strain, $e_D=e_N-e_V$. The volumetric strain is computed by the volume variation of the Delaunay tetrahedra as $e_V= \Delta V/ 3V_0$ and is assumed to be the same for all facets belonging to a given tetrahedron. Beyond the elastic limit, $-\sigma_{bc}$ models pore collapse as a linear evolution of stress for increasing volumetric strain with stiffness $H_{c}$ for $-e_V \leq e_{c1} = \kappa_{c0} e_{c0}$: $\sigma_{bc} = \sigma_{c0} + \langle-e_V-e_{c0}\rangle H_c(r_{DV})$; $H_c(r_{DV})=H_{c0}/(1 + \kappa_{c2} \left\langle r_{DV} - \kappa_{c1} \right\rangle)$; $\sigma_{c0}$ is the mesoscale compressive yield stress; $r_{DV}=e_D/e_V$ and $\kappa_{c1}$, $\kappa_{c2}$ are material parameters. Compaction and rehardening occur beyond pore collapse ($-e_V \geq e_{c1}$). In this case one has $\sigma_{bc} = \sigma_{c1}(r_{DV})$ $\exp \left[( -e_{V}-e_{c1} ) H_c(r_{DV})/\sigma_{c1}(r_{DV}) \right]$ and $\sigma_{c1}(r_{DV}) = \sigma_{c0} + (e_{c1}-e_{c0}) H_c(r_{DV})$. 

\textbf{Friction due to compression-shear.} The incremental shear stresses are computed as  $\dot{t}_M=E_T(\dot{e}^*_M-\dot{e}^{*p}_M)$ and \mbox{$\dot{t}_L=E_T(\dot{e}^*_L-\dot{e}^{*p}_L)$}, where  \mbox{$\dot{e}_M^{*p}=\dot{\xi} \partial \varphi / \partial t_M$}, \mbox{$\dot{e}_L^{*p}=\dot{\xi} \partial \varphi / \partial t_L$}, and $\xi$ is the plastic multiplier with loading-unloading conditions  $\varphi \dot{\xi} \leq 0$ and $\dot{\xi} \geq 0$. The plastic potential is defined as \mbox{$\varphi=\sqrt{t_M^2+t_L^2} - \sigma_{bs}(t_N)$}, where the nonlinear frictional law for the shear strength is assumed to be $\sigma_{bs} = \sigma_s + (\mu_0 - \mu_\infty)\sigma_{N0}[1 - \exp(t_N / \sigma_{N0})] - \mu_\infty t_N$; $\sigma_{N0}$ is the transitional normal stress; $\mu_0$ and $\mu_\infty$ are the initial and final internal friction coefficients.  

Each material property in LDPM governs part of the concrete behavior under loading. The normal elastic modulus, which refers to the stiffness for the normal facet behavior, $E_0$, governs the LDPM response in the elastic regime, along with the coupling parameter $\alpha$. Approximately, the macro scale Young's modulus $E$ and Poisson's ratios $\nu$ can be calculated as $E = E_0(2+3\alpha)/(4+\alpha)$ and $\nu = (1-\alpha)/(4+\alpha)$. Typical concrete Poisson's ratio of about 0.18 is obtained by setting $\alpha$ = 0.25 \cite{Cusatis2011b}. 

The tensile strength, $\sigma_t$, and characteristic length, $\ell_{t}$, together define the softening behavior due to fracture in tension of LDPM facets \cite{Cusatis2011b}. The mesoscale fracture energy can be obtained by the relation $G_t = \ell_{t} \sigma_t^2 /2E_0$. Calibration of $\sigma_t$ and $\ell_t$ is typically achieved by fitting fracture-related experimental data, e.g. the load-displacement curve of a notched three-point-bending test.

The softening exponent, $n_t$, governs the interaction between shear and tensile behavior during softening at the facet level. One obtains more ductile behavior in both compression and tension by increasing $n_t$, however the increase is more pronounced in compression than in tension. The shear strength, $\sigma_s$, is the facet strength for pure shear and affects the macroscopic behavior in unconfined compression. Compressive yielding stress, $\sigma_{c0}$; initial hardening modulus, $H_{c0}$; transitional strain ratio, $k_{c0}$; triaxial parameters, $k_{c1}$ and $k_{c2}$; and densified normal modulus, $E_d$, define the behavior of the facet normal component under compression and affect mostly the macroscopic behavior in hydrostatic and highly-confined compression. The initial internal friction, $\mu_0$; and transitional stress, $\sigma_{N0}$, mainly govern the mechanical response in compression and have no influence on the tensile behavior in LDPM. At the macroscopic level they mostly affect the compressive behavior at zero or low confinement. Detailed descriptions of effect and function of other LDPM mesoscale parameters can be found in Cusatis et al. \cite{Cusatis2011b}.
 
Besides the compatibility and constitutive equations discussed above, the governing equations of the LDPM framework are completed through the equilibrium equations of each individual particle. 

LDPM has been utilized successfully to simulate concrete behavior under various loading conditions \cite{Cusatis2011a,Cusatis2011b,Wan2016MC, WanUHPCc, WanPhDNU}. Furthermore, it was recently extended to properly account for fiber reinforcement \cite{Schauffert2012I,Schauffert2012II} and it has the ability to simulate the ballistic behavior of ultra-high performance concrete (UHPC) \cite{Smith2014}. In addition, LDPM showed success in structural scale analysis using multiscale methods \cite{HOMO,CG,Euro-C, WanConcreep2015, WanConcreep2015b, WanKEM2016}.

\subsection{Aging Formulation}

Experimental studies show that elastic modulus (and strength) evolution at early age depends not only on the degree of chemical reactions, but also, independently, on the curing temperature history \cite{Kim1998, Cervera2000, Kim2002I}. To account for this effect the aging degree $\lambda$ is introduced:  


\begin{equation}
\label{ad}
\dot{\lambda} = 
\begin{cases} 0 & \mathrm{for} \, \alpha < \alpha_0 \\
\left(\frac{T_{max} - T}{T_{max} - T_{ref}}\right)^{n_\lambda}(B_{\lambda} - 2A_{\lambda}\alpha)\dot{\alpha} & \mathrm{for} \, \alpha \ge \alpha_0
\end{cases}
\end{equation}
where $T_{max}$ = 120~$^\circ$C, $T_{ref}$  = 22~$^\circ$C, $B_{\lambda} = [1 + A_{\lambda}(\alpha^2_\infty - \alpha^2_0)]/(\alpha_\infty - \alpha_0)$, $\alpha_\infty$ = 1, $\alpha_0$ = total reaction degree at setting, $n_{\lambda}$ and $A_{\lambda}$ are model parameters obtained from fitting experimental data related to aging mechanical properties, and $\alpha$ is the total degree of reaction as previously introduced. 

At constant curing temperature, experiments show that elastic modulus and strength are monotonously increasing functions of concrete age \cite{Karte2015,Chamrova2010,Schutter1996}. Hence, the aging degree must also be a monotonically increasing function of the total reaction degree. This can be enforced by imposing the following restriction on the parameter $A_\lambda$: $A_\lambda  < A_\lambda^{max}=(2\alpha_\infty-2\alpha_0-\alpha^2_\infty+\alpha^2_0)^{-1}$. 
Furthermore, it is worth pointing out that with typical values (0.1-0.3) of the hydration degree at setting, $\alpha_0$, for $A_\lambda<0.25$ and constant temperature, the function $\lambda(\alpha)$ is practically coincident with a straight line.  

The aging degree can be calibrated either with elastic modulus data or with strength data relevant to various concrete ages and at least two different temperatures. In this study elastic modulus data are used and, consequently, the normal modulus, $E_0$, is assumed to have a linear relation with aging degree $\lambda$: 

\begin{equation} 
\label{age1}
E_0 = E^{\infty}_0 \lambda 
\end{equation}

Tensile strength, $\sigma_t$, compressive yielding stress, $\sigma_c$, and transitional stress, $\sigma_{N0}$, on the other hand, are assumed to have power-law type relations with aging degree:

 \begin{equation}
\sigma_t = \sigma_t^{\infty} \lambda^{n_t} ; \:
\sigma_c = \sigma_c^{\infty} \lambda^{n_c} ; \:
\sigma_{N0} = \sigma_{N0}^{\infty} \lambda^{n_{N0}} 
\end{equation}
 
where $n_t$, $n_c$, $n_{N0}$ are positive constants. For simplicity and without sacrificing the quality of the simulations the power-law exponents can be equated, introducing a single aging exponent $n_a = n_t = n_c = n_{N0}$. Lastly, the tensile characteristic length, $\ell_{t}$, is assumed to be a linear decreasing function of aging degree. This dependence is introduced to describe the well known brittleness increase with age:

 \begin{equation}
 \label{age4}
\ell_{t} = \ell^{\infty}_{t} (k_a(1-\lambda) + 1)
\end{equation}

where $k_a$ is a positive constant. All the aging functions are formulated such that the corresponding parameters approach their asymptotic values for $\lambda$ approaching the value of 1. The other LDPM mesoscale parameters, are assumed age-independent due to a lack of relevant experimental data on the response in compression under confinement.
 



\section{Experimental Characterization of Early Age Behavior of UHPC}\no

In order to calibrate and validate the proposed early age model, an experimental campaign was carried out to characterize the early age mechanical behavior of a UHPC material. The mixture proportions for the adopted mix design are reported in Table~1. The material composition consists of LaFarge Type~H cement, F-50 Ottawa sand, Sil-co-sil 75 silica flour, Elkem ES-900W silica fume, ADVA-190 Superplasticizer and tap water. The maximum particle size, 0.6 mm, is limited to that of silica sand, which is a foundry grade Ottawa sand  \cite{CorTuf2009}.

\begin{table}[htbp]
\centering
\begin{tabular}{l l c c}
\multicolumn{2}{c}{Table 1. Constituents and Mixing Proportions of UHPC CorTuf}\\
\hline
Ingredient	&Type	&Proportion	&Weight per kg	\\
\hline
Cement 	&Lafarge Type H&1.0000	&0.3497	\\
Sand 	&F-50&0.9674	&0.3383	\\
Silica Flour  &	Sil-co-sil 75&0.2768	&0.0968	\\
Silica Fumes 	&Elkem ES-900W&	0.3890	&0.1360	\\
Superplasticizer&	ADVA-190&	0.0180&	0.0063\\
Water	&Tap Water&	0.2082&	0.0728	\\
\hline
\end{tabular}
\end{table}

The baseline curing regime for the adopted UHPC \cite{CorTuf2009} consisted of casting on day 0, demolding on day 1, curing in a 100\% humidity room (HR) with elevated room temperature of approximately 28~$^\circ$C until day 7, followed by curing for 4 days in 85~$^\circ$C water bath (WB), and 2 days drying in the oven at 85~$^\circ$C. Preliminary unconfined compression tests using 50.8$\times$101.6 mm (2$\times$4 in) cylinders were carried out at nominal ages of 1 day, 7 days, 11 days and 13 days. At least 3 specimens were tested for each age to obtain mean and standard deviation of material properties. Measured compressive strengths for the 4 different ages were 13~MPa $\pm$ 31$\%$, 66~MPa $\pm$ 23$\%$, 123~MPa $\pm$ 29$\%$, and 120~MPa $\pm$ 21$\%$. As seen from the results, oven curing does not provide the expected increase in compressive strength. Rather, a very slight (-2$\%$) decrease in the average compressive strength was observed. Considering the number of tested specimens (3) and the amount of scatter, no statistically significant change in mean value was detected. Consequently, later experiments disregarded the oven curing procedure.

To study the effects of hot water bath curing on strength gain, two curing protocols with and without hot water bath curing were explored. A first group of specimens was kept in the humidity room (HR) for 14 days, a second group, instead, was kept in the humidity room for 7 days after which it was placed in hot water bath (WB) at 85~$^\circ$C for another 7~days. Both groups were later exposed to the same laboratory conditions (about 22~$^\circ$C and 50\%~RH). Unconfined compression tests using 50.8$\times$50.8 mm (2$\times$2 in) cylinders, three-point-bending (TPB) tests with half-depth notched 25.4$\times$25.4$\times$127~mm (1$\times$1$\times$5 in) beams, and tensile splitting tests using 76.2$\times$25.4~mm (3$\times$1 in) disks were carried out on nominal ages of 3, 7, 14, and 28 days. Circumferential expansion control for compression tests and extensometer control for three-point-bending and splitting tests were utilized to avoid brittle failure and to obtain full post-peak information. Furthermore, a friction reduction layer (moly dry lubricant) was applied to the load platens for all compression tests. \\

\begin{figure} [H]
\centering
\includegraphics[height=1.8in,valign=t]{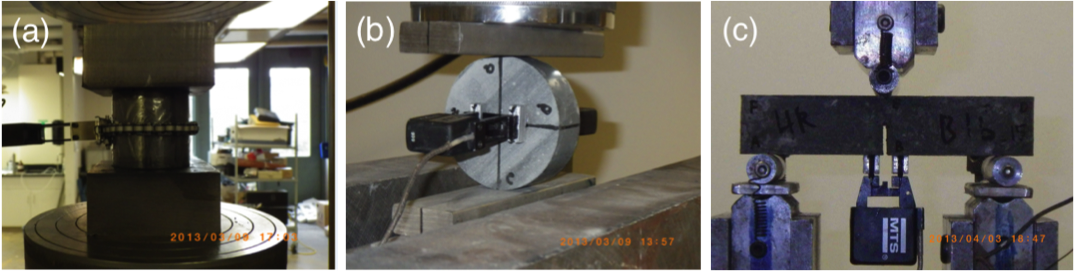}
\caption{UHPC experiment setup: (a) cylinder compression, (b) splitting, (c) beam 3-point-bending}
\label{exp}
\end{figure}

\subsection{Unconfined Cylinder Compression Test}

Unconfined compression tests were performed in a closed loop servo-hydraulic MTS load frame with a maximum capacity of 4.4~MN (1000~kips). A circumferential extensometer was utilized to control expansion of the specimens with the goal to obtain post-peak response (see Fig.~\ref{exp}a). After centering the specimen, a preload of 1-5\% of the expected peak was applied before the actual test commenced. 

For compression tests, two phases of control were used: first stroke control followed by circumferential expansion control. Both control modes applied a loading rate of 0.001~mm/s. The mode of control was switched at estimated 30$\%$ of the potential peak load. Compression test results can be found in Table~2 and Fig.~\ref{compression}, where stress is calculated as $\sigma = P/A$ (load/area) and strain as $\epsilon = \Delta d/H$ (relative displacement between loading platens/specimen height). Utilized for the calculations of stress and strain for each specimen, the measurements of height $H$ and diameter $D$ with standard deviations are also shown in the figures. 

As Table~2 shows, at 14 days, samples that were cured in hot water bath exhibited about 28\% higher strength compared with the steam curing. However, after 14 more days in air (from day 14 to day 28) the strength gain was almost completely lost and the water cured samples were on average only 1\% stronger than the others. This effect was, most likely, the result of both cooling and shrinkage cracking. 

\begin{table}[htbp]
\centering
\begin{tabular}{c c c c c}
\multicolumn{5}{l}{Table 2. Unconfined Cylinder Compression Test Results}\\
\hline
Nominal Age [d]&Actual Ave. Age [d]& Curing &Strength [MPa]&No. of Specimens\\
\hline
3 	&2.9& 	HR &	53.8 	$\pm$ 21.7\% &4 \\
7 	&7.0 	&HR &	78.4 $\pm$ 10.6\% & 	4 \\
14 &	14.5& 	HR &	99.3 $\pm$ 15.6\% &	4 \\
28 	&29.8 	&HR &	115.6 $\pm$2.2\% 	&3 \\
14 	&14.4 &	HR + WB &	127.2 $\pm$ 14.6\% &	4 \\
28 &	30.0 	&HR + WB 	&116.7 $\pm$ 16.4\%&	5 \\

\hline
\end{tabular}
\end{table}

\subsection{Brazilian Test}

Indirect tensile splitting (Brazilian) tests were performed in a MTS load frame with a maximum axial capacity of about 1~MN (220,000~lbs). The testing procedure was developed according to ASTM~C496. Loading blocks with dimensions of 15$\times$38$\times$159~mm (0.6$\times$1.5$\times$6.25~in) were used. Initially, localized failure in compression was avoided by the utilization of wooden support strips. However, their usage introduced an unwanted compliance at the beginning of the test. Therefore, aging tests had been ran without wooden strips. Contrary to ASTM C496 which calls for standard cylinder dimensions of e.g. 50.8$\times$101.6 mm (2$\times$4 in), cylindrical disks of 76.2 mm (3 in) in diameter and 25.4 mm (1 in) in thickness were tested (see Fig.~\ref{exp}b). This specimen geometry allows a relatively more stable transition to the post-peak regime. 

After the specimen was placed firmly in the MTS Load Frame and preloaded with approximately 1$\%$ of estimated peak to ensure contact between specimen and loading supports, the first phase of the actual test was carried out in load control with a loading rate of 0.1~kN/sec. As soon as 50$\%$ of the estimated peak was reached, the test was switched to displacement control to save experiment time and avoid sudden brittle failure. The displacement was taken by means of a transducer mounted on the front face of the samples (see Fig.~\ref{exp}b). A relatively stable displacement rate was found to be 5$\times$10$^{-7}$~mm/sec for the highly brittle UHPC specimens. After reaching a stable post-peak response in displacement control, the opening rate was accelerated to save testing time while ensuring stability. At least 3 specimens were available for testing on each age. In spite of all efforts and great precautions only for about half of the young specimens post-peak response could be obtained, and for none of the higher curing degrees. The Brazilian test results, including age of concrete at testing, curing conditions, nominal strength, number of specimens and dimension measurements, are shown in Table~3. The nominal tensile splitting strength is computed as $\sigma_u = 2P/\pi hd$ ([2$\times$load]/[$\pi$$\times$height$\times$diameter]) with the actual dimensions for each specimen. The obtained tensile strength values do not follow the expected aging trend. The most likely reason can be found in the almost futile attempt to obtain post-peak data by modifying the loading rate during the tests which might have influenced the observed peak values. 

\begin{table}[htbp]
\centering
\begin{tabular}{p{1.5cm}p{2cm}p{2cm}p{3cm}p{2cm}p{2.8cm}p{2.5cm}}
\multicolumn{7}{l}{Table 3.  \color{black} Brazilian Test Results}\\
\hline
Nominal Age &	Actual Ave. Age &Curing &Splitting Tensile Strength [MPa] &Number of Specimens & Diameter~d~[mm] & Height~h~[mm]\\	
\hline
3 &3.0&	HR &	4.5 $\pm$ 10.4\%	&3  & 76.3 $\pm$ 0.3\% & 23.2 $\pm$ 9.0\%\\
7 	&7.0	&HR &	5.2 $\pm$ 9.1\% &	3 & 76.3 $\pm$  0.2\% & 24.5 $\pm$ 4.0\% \\
14 &	14.9	&HR &	5.5 $\pm$ 0\%	&1 & 76.4 $\pm$ 0.0\% & 23.3$\pm$ 0.0\%\\
28 &	29.6	&HR &	4.6 $\pm$	17.1\% &3 & 76.2 $\pm$ 0.3\% & 24.5$\pm$ 4.5\%\\
14 &	14.6	&HR + WB &	11.4 $\pm$ 22.7\%	&3 & 76.3 $\pm$ 0.3\% & 25.4 $\pm$ 10.4 \%\\
28 &	29.2	&HR + WB 	&9.4	$\pm$ 12.2\% &3 & 76.4$\pm$ 0.2\% & 24.7 $\pm$ 3.3\%\\
\hline
\end{tabular}
\end{table}
\color{black}

\subsection{Three-point-bending Fracture Test}

Beam specimens with 25.4$\times$25.4$\times$127~mm (1$\times$1$\times$5~in) dimensions were cast for TPB tests (see Fig.~\ref{exp}c) with notches of 50\% relative depth. The nominal span (distance between bottom supports) was 101.6~mm (4~in). An extensometer sensor was glued to the bottom of the specimen with the notch between its two feet. After a pre-load of up to 5$\%$ of the expected peak was applied, the specimen was loaded in crack mouth opening (CMOD) control with an initial loading rate of 0.0001~mm/sec. This rate was increased in the late post-peak phase to save total testing time while ensuring a fully recorded softening behavior. The test results and nominal stress-strain curves can be found in Table~4 and Fig.~\ref{bending}, where the nominal flexural stress is obtained by equation $\sigma = 3PS/2BH^2$ ([3$\times$load$\times$test span]/[2$\times$specimen width$\times$specimen depth$^2$]) and nominal strain by $\epsilon$ = CMOD/specimen depth, the total fracture energy $G_F$ is calculated as the area under the force-displacement curve divided by the ligament area. Geometry measurement statistics, which are used to calculate $\sigma$, $\epsilon$, and $G_F$, are also included in the figures. 

\begin{table}[htbp]
\centering
\begin{tabular}{c c c c c}
\multicolumn{5}{l}{Table 4. Three-point-bending Test Results}\\
\hline
Nominal Age [d] &	Actual Ave. Age [d] &	Curing &	Max. Nominal Stress [MPa]&No. of Specimens\\
\hline
3 	&2.9 &	HR& 	1.79 $\pm$ 4.3\%		&3\\
7 	&7.0 	&HR &	2.12 $\pm$ 8.4\%		&3\\
14 &	14.0 &	HR &	2.65	$\pm$ 10.2\%	&4\\
28 &	28.0& 	HR &	2.55	$\pm$ 7.3\% &	4\\ 
14 	&14.4 &	HR + WB& 	3.65 $\pm$ 2.5\% &		4\\
28 &	28.0 &	HR + WB &	3.10	$\pm$7.8\% &	4\\
\hline
\end{tabular}
\end{table}
\color{black}

In general, all experiments show the expected increase in strength with age. This is also observed for the TPB tests. However, also a strength decrease from 14~days to 28~days was recorded for all test types and curing conditions, except compression with humidity room curing. This surprising result defies theory and, at the present, can only be explained by drying shrinkage damage, affecting those specimens that were stored in air after the hot water bath.


\subsection{Relative Humidity Measurements}

A quintessential input for any mechanical analysis utilizing the proposed aging framework are time-dependent spatial fields of reaction degrees. These are a ``by-product'' of the coupled chemo-hygro-thermal analysis using the HTC model. The two essential fields of the multi-physics problem are relative humidity $h$ and temperature $T$. Consequently, in order to calibrate the HTC model for the specific UHPC of the case study, $h$ and$T$ measurement data are required on top of literature values for the reaction kinetics and diffusion related parameters. In this investigation RH measurements at the center of the specimens as well as at the specimen boundaries (ambient environment) were conducted with three curing routines: 14 days HR, 7 days HR + 7 days WB, and fully sealed (self desiccation) at room temperature. For each curing method, five 50.8$\times$50.8 mm (2$\times$2 in) cylinders were measured by RH\&T sensors, see Fig.~\ref{exprh}. At the time of casting, a straw with one side closed by nylon mesh and housing a RH\&T sensor, was vertically inserted into each specimen with the perforated end centered in the specimen. The other end of the straw protruding beyond the specimen surface was sealed with moisture-tight silicone sealant. The RH\&T sensors remained for the whole measurement period in the self-desiccation specimens, which were sealed with plastic molds (Fig.~\ref{exprh}c), silicone sealant and plastic films. For the unsealed samples, during and after 100\% RH humidity room curing, the RH\&T sensors were taken out and put back in about once a day to avoid condensation while the operation time was minimized to preserve accurate RH measurements. Utilizing straws instead of embedding sensors directly in the specimen was motivated by the intent to 1) protect the sensors, 2) be able to reuse the sensors, and 3) easily check and replace sensors in case of malfunction. The temperature measurement is not reported because no appreciable hydration related temperature increase was observed, owing to the small dimension of the specimens and the relatively high thermal conductivity of concrete. A time history of RH at the center of the specimens was obtained for each curing routine, as shown in Fig.~\ref{seal} $\sim$ \ref{hrwb}. For the sealed specimens, the RH level at the center dropped relatively rapidly and reached about 69\% on day 40, which is much lower than that of normal concrete and high performance concrete \cite{Yang2004, Zhang2012}. This is due to the very low water/cement ratio and the low permeability of the UHPC (the water/cement ratio for this UHPC is only 0.2). The amount of free water decreases relatively fast as water is being consumed by cement hydration and pozzolanic reactions. Overall, the three parallel experiments of each curing regime show very consistent RH evolutions, as shown in Figs.~\ref{seal}(a)~$\sim$~\ref{hrwb}(a).  

\begin{figure}[H]
\centering
\includegraphics[height=1.3in]{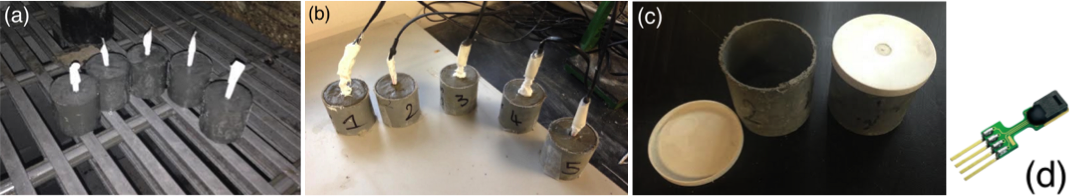}
\caption{Relative humidity measurement: (a) specimens curing, (b) measurement setup, (c) plastic molds, (d) RH\&T sensor}
\label{exprh}
\end{figure}

\section{LDPM Modeling of Early Age UHPC Behavior: Calibration and Validation}

 \subsection{HTC Model Calibration and Validation}
 
The actual HTC model calibration is an inverse analysis task. For that purpose the aforementioned 50.8$\times$50.8~mm (2$\times$2~in) cylinder specimens of RH measurements are simulated. The HTC input data include boundary conditions and material properties. The environmental boundary conditions are as follows: for HR curing: 28$^\circ$C and 100\% RH for 14 days then 50\% RH afterwards; for HR+WB curing: 28$^\circ$C and 100\% RH for 7 days, then 85$^\circ$C and 100\% RH for the next 7 days and afterwards 50\% RH at room temperature; and for self desiccation curing: RH sealing and room temperature. The material properties and HTC model parameters used in the simulations are listed in Table~5, in which the data source is also reported.
The internal RH measurements under sealed conditions are in particular used for the calibration of the self desiccation parameters. The moisture diffusion parameters are calibrated by humidity room curing RH measurements of 50 days and self desiccation RH measurements of 40 days, starting from casting, and validated by HR+WB curing RH measurements of 62 days of data, see Figs.~\ref{seal}~$\sim$~\ref{hrwb}, which include also the environmental RH. 
The simulated RH evolution agrees very well with those of the experimental investigations for the applied curing regimes. Figs.~\ref{seal}(b)~$\sim$~\ref{hrwb}(b) show the evolutions of cement hydration degree, silica fume (SF) reaction degree, and total reaction degree. The total reaction degree of HR curing increases gradually and reaches about 0.89 on day 60, while the aging degree under HR+WB curing reaches its asymptote of 1.0, fairly early, at around 14 days of age (0.997 to be exact). The thermal activation results in a higher total reaction degree which also evolves with a higher rate. Figs.~\ref{htcrh} and \ref{htcad} present the spatial distributions of relative humidity and total reaction degree, respectively, after 3, 14, and 28 days of HR curing (14 days in 100\% humidity room with elevated room temperature of 28 $^\circ$C). The relative humidity exhibits relatively high spatial variability, especially as the boundary condition changes at the early ages. On the contrary, the total reaction degree shows low spatial variability of less than about 1\%, due to the small dimension of the tested laboratory scale specimens.

\begin{table}[htbp]
\centering
\begin{tabular}{l l l l}
\multicolumn{3}{l}{Table 5: HTC Parameters}\\
\hline
Material property name [unit] &  Symbol & Value & Source\\
\hline
Density [kg/m$^3$] & $\rho$ & 2400 & this paper \\
Isobaric heat capacity [J/kg$^\circ$C] & $c_t$ & 1100  & \cite{DiLuzio2009II} \\
Heat conductivity [W/m$^\circ$C] & $\lambda_t$ & 5.4 & \cite{DiLuzio2013} \\
Cement hydration enthalpy [kJ/kg] & $\tilde{Q}_c^{\infty}$ & 500 &\cite{DiLuzio2009II} \\
Silica fume reaction enthalpy [kJ/kg] & $\tilde{Q}_s^{\infty}$  & 780 &\cite{DiLuzio2009II} \\
Hydration activation energy/R [K] & $E_{ac}/R$ & 5490 & \cite{Cervera1999} \\
Silica fume reaction activation energy/R [K] & $E_{as}/R$ & 9700 &\cite{DiLuzio2009II} \\
Diffusivity activation energy/R [K] & $E_{ad}/R$ & 2700 &\cite{DiLuzio2009II} \\
Silica fume efficiency [-] & SF$^{eff}$ & 0.9 &\cite{DiLuzio2009II} \\
Polymerization activation energy/R [K] & $E_{ap}/R$ & 6000 &\cite{DiLuzio2009II} \\
Hydration parameter [h$^{-1}$] & $A_{c1}$ & 2 $\times$ 10$^8$ &\cite{DiLuzio2009II} \\
Hydration parameter [-] & $A_{c2}$ & 1 $\times$ 10$^{-6}$ &\cite{DiLuzio2009II} \\
Hydration parameter [-] & $\eta_c$ & 6.5 &\cite{DiLuzio2009II} \\
Silica fume reaction parameter [h$^{-1}$] & $A_{s1}$ & 5 $\times$ 10$^{14}$ &\cite{DiLuzio2009II}  \\
Silica fume reaction parameter [-] & $A_{s2}$ & 1 $\times$ 10$^{-6}$ &\cite{DiLuzio2009II} \\
Silica fume reaction parameter [-] & $\eta_s$ & 9.5 &\cite{DiLuzio2009II} \\
Moisture diffusion parameter [mm$^2$/h] & $D_0/c$ & 1 $\times$ 10$^{-4}$ & this paper \\
Moisture diffusion parameter [mm$^2$/h] & $D_1/c$ & 3.1 & this paper  \\
Moisture diffusion parameter [-] & n & 3.9 & this paper \\
Self desiccation parameter [-] & $g_1$ & 1.5 & this paper \\
Self desiccation parameter [-] & $k_{vg}^c$ & 0.2 & this paper \\
Self desiccation parameter [-] & $k_{vg}^s$ & 0.4 & this paper \\
\hline
\end{tabular}
\end{table}

\begin{figure} [H]
    \centering
     \includegraphics[width=7.5in,valign=t]{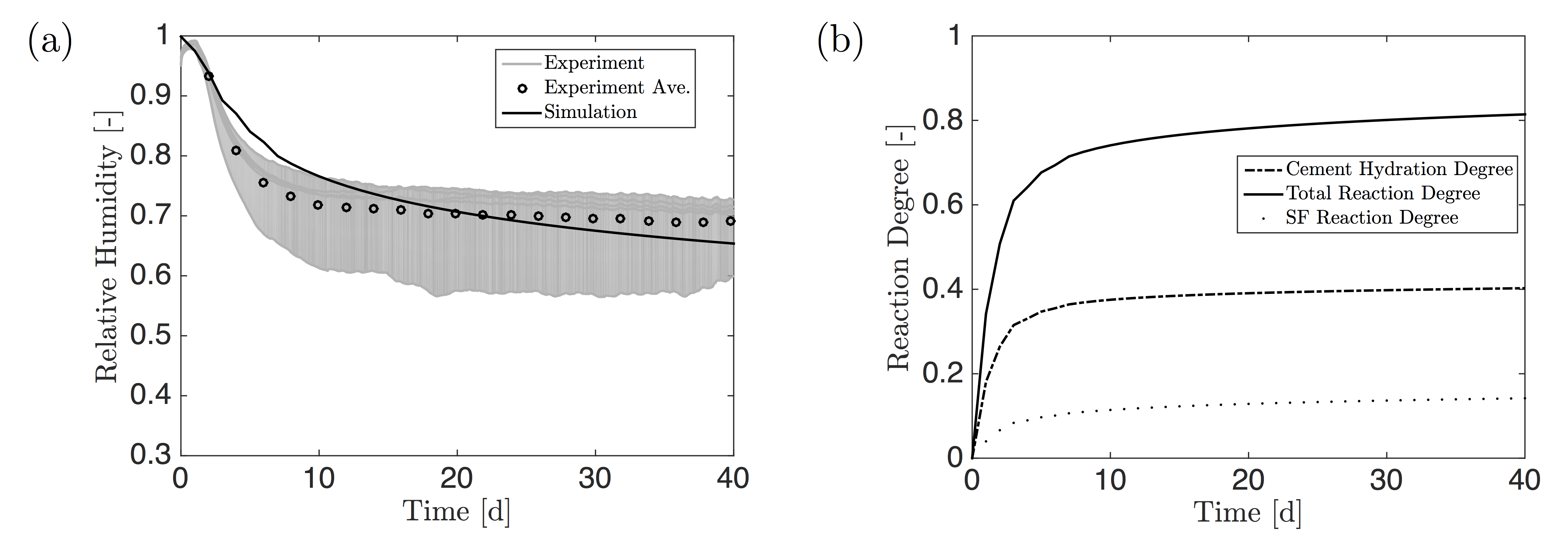} 
    \caption{HTC calibration of sealed specimens for self desiccation study: (a) RH evolution and (b) simulated reaction degrees}
    \label{seal}
\end{figure}

\begin{figure} [H]
    \centering
   \includegraphics[width=7.5in,valign=t]{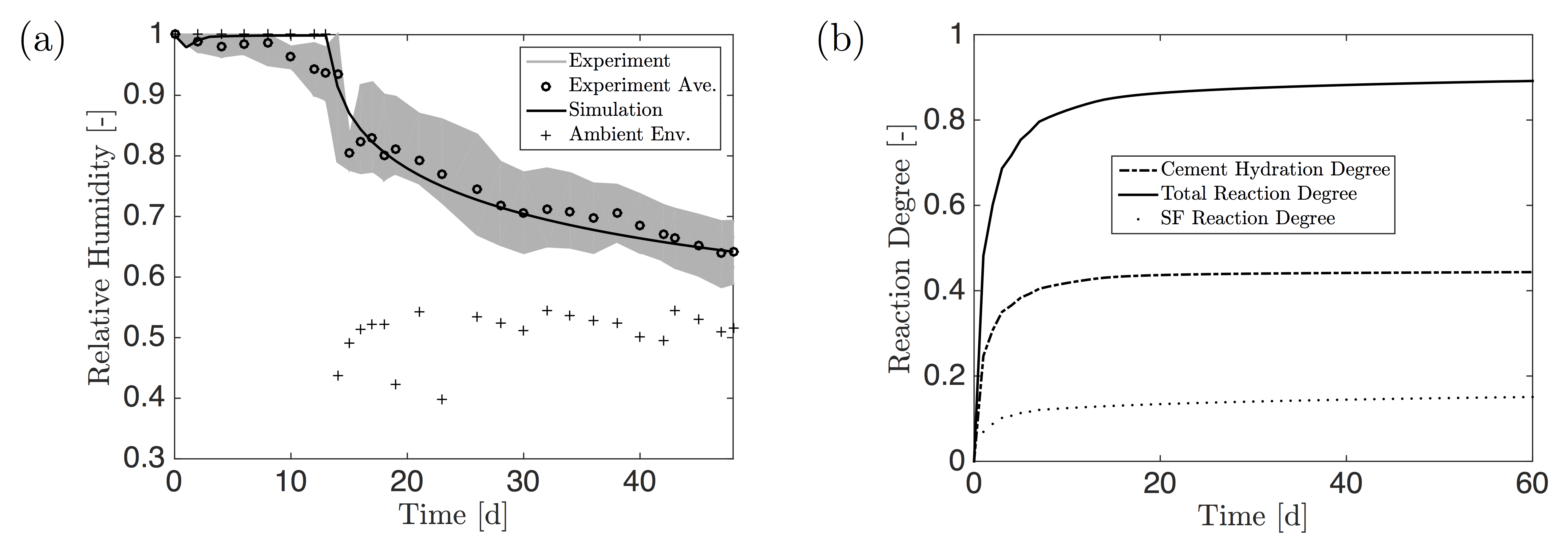} 
    \caption{HTC calibration by specimens under HR curing: (a) RH evolution and (b) simulated reaction degrees}
    \label{hr}
\end{figure}

\begin{figure} [H]
    \centering
    \includegraphics[width=7.5in,valign=t]{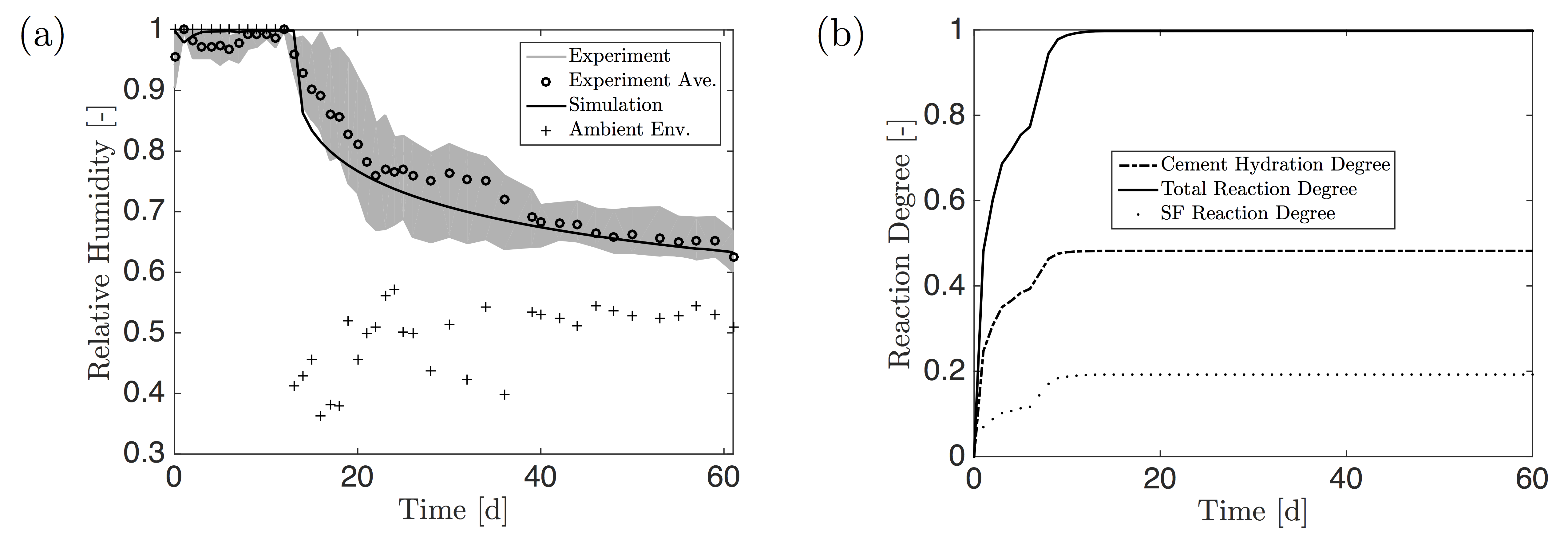} 
    \caption{HTC validation using specimens under HR + WB curing: (a) RH evolution and (b) simulated reaction degrees}
    \label{hrwb}
\end{figure}

\begin{figure} [H]
    \centering
     \includegraphics[width=7.5in]{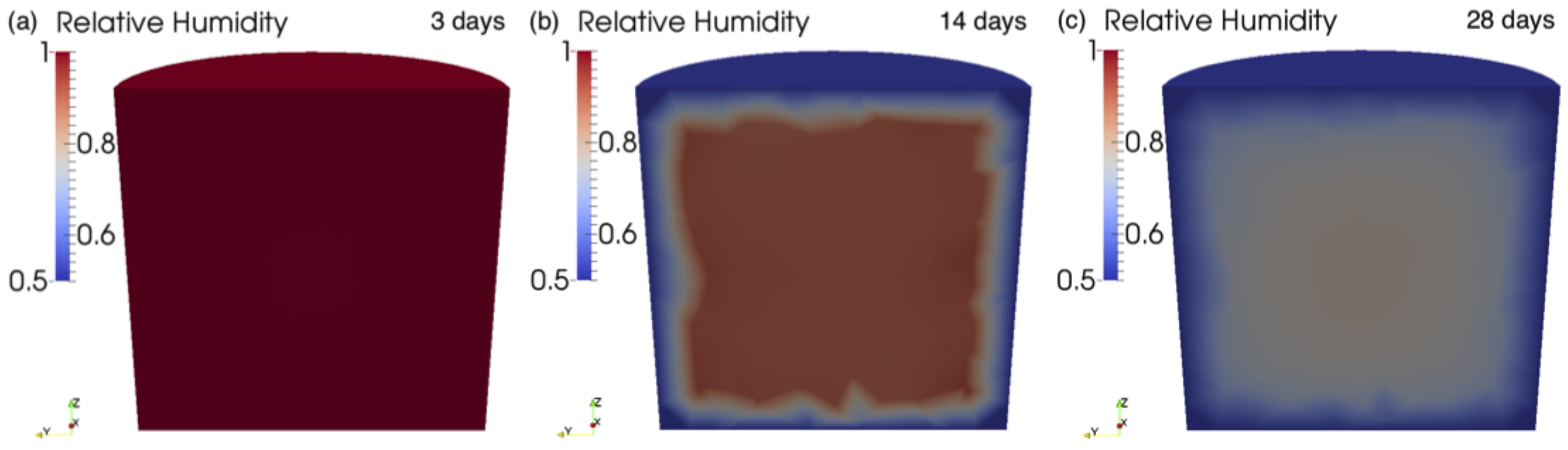}
        \caption{Relative humidity from HTC simulations under HR curing on (a) 3, (b) 14, and (c) 28 days of age}
        \label{htcrh}
\end{figure}

\begin{figure} [H]
    \centering
    \includegraphics[width=7.5in]{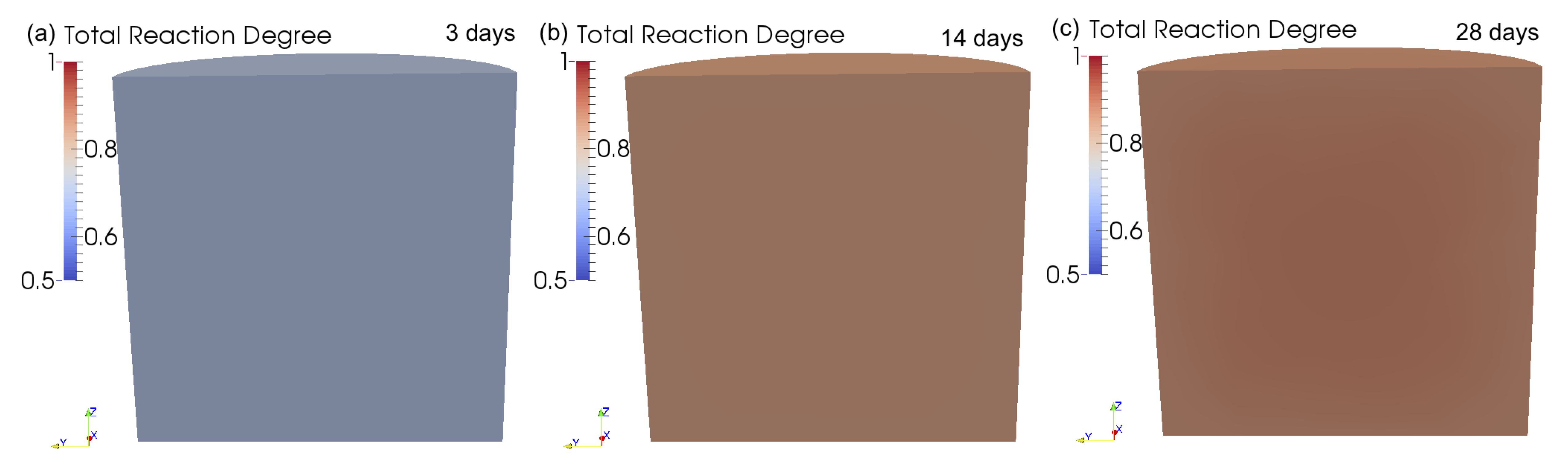}
    \caption{Total reaction degree from HTC simulations under HR curing on (a) 3, (b) 14, and (c) 28 days of age}
    \label{htcad}
\end{figure}

\subsection{Calibration of LDPM - Aging Degree and Aging Functions}

The first step in the calibration of the aging model is the calibration of the aging degree that, in this study and consistently with previous work of the last author \cite{DiLuzio2013}, is pursued on the basis of elastic modulus measurements. Identified values of the LDPM normal modulus for the case of humidity room curing (HR, $T=28 ^\circ$C) are plotted in Fig.~\ref{trdE}a as function of the total reaction degree. As shown in Fig.~\ref{trdE}a, the best fit of the data points can be obtained by setting $E_0^\infty=75$ GPa, $\alpha_0=0.278$, and $A_\lambda=0.2$. The parameter $n_\lambda$ is then calibrated by fitting the modulus data relevant to the water bath curing at 14 days. This gives $n_\lambda=0.1$. Finally, the parameters $n_a$ and $k_a$ are identified by fitting strength and fracture data. The optimum fit is obtained by setting $n_a=2.33$ and $k_a=22.2$. It is worth observing that the adopted aging functions require that $E/E_0^\infty = (\sigma_t/\sigma_t^\infty)^{1/n_a} \approx \sqrt{\sigma_t/\sigma_t^\infty}$ which is the typical relationship between elastic modulus and strength as vastly adopted in the literature \cite{ACI2008}. The resulting aging degree evolutions for WB and HR curing regimes are shown in Fig.~\ref{trdE}b. Similarly to the aforementioned total reaction degree, the aging degree with WB curing reaches its asymptotic value, fairly early, at around 14 days of age (0.95), while the aging degree of HR curing increases gradually and reaches about 0.84 at 60 days of age. The thermal activation results in a higher aging degree which also evolves at a higher rate. 


\begin{figure} [H]
\centering
\includegraphics[width=7in]{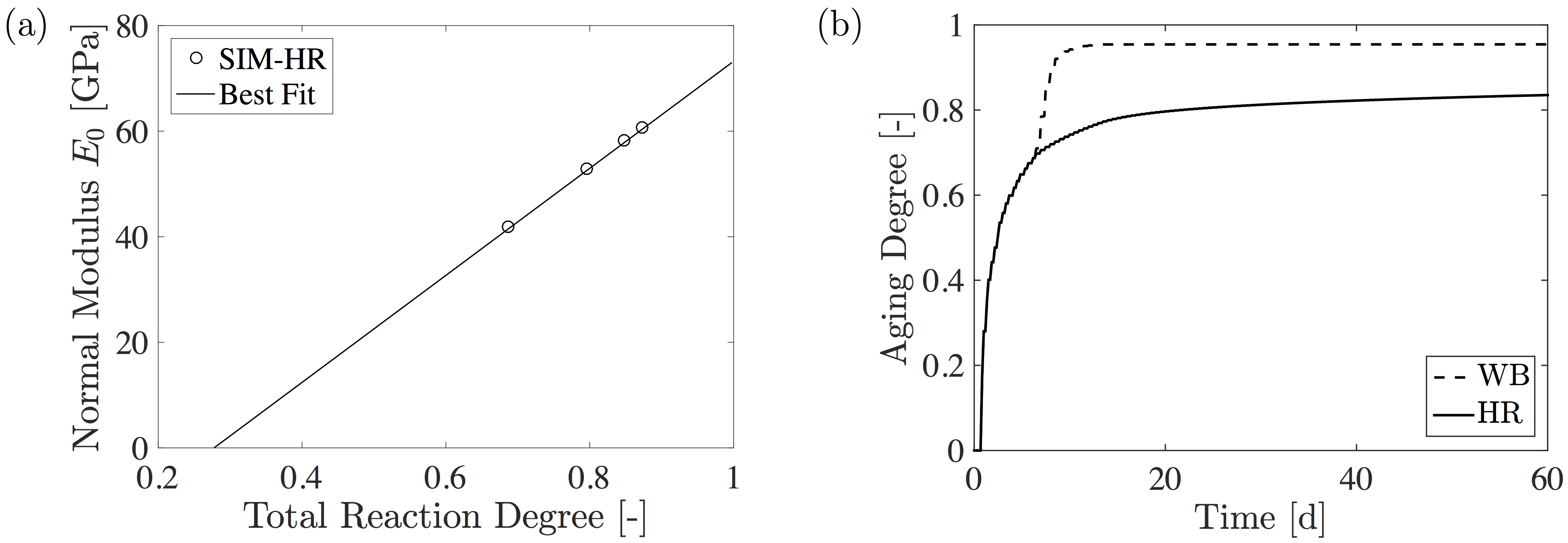}
\caption{(a) LDPM normal modulus vs. total reaction degree for humidity room curing (b) Aging degree evolution for water bath curing and humidity room curing}
\label{trdE}
\end{figure}

%

All the LDPM simulations utilize a coarse-grained aggregate size in the range 2$\sim$4~mm. In Cusatis et. al. \cite{Cusatis2011a, Cusatis2011b}, it was demonstrated that, although approximated, the coarse-grained simulations provide good accuracy with a significant reduction in computational time. A complete list of all mesoscale parameters used in LDPM for all ages investigated can be viewed in Table~6, where the second column lists the asymptotic values. HR stands for 14 days in humidity room curing, WB stands for 7 days humidity room + 7 days water bath curing regime, after which for both curing routines the specimens are stored at room temperature with RH around 50\%. Due to the small specimen dimensions, the spatial variability of aging degree for all investigated ages is below about 1\%, see the standard deviation values for aging degree shown in Table~6. Hence, the difference in the simulation results including and excluding spatial variability is negligible for the investigated case, see Fig.~\ref{marsHTC}. Thus, spatially averaged aging degree and corresponding material properties are utilized for the simulated compression, splitting and bending tests. Also because the cylinders, cylindrical disks and beam specimens used in compression, Brazilian, and bending tests have similar size, the same material properties on corresponding ages are used for all the specimen types. 

To ensure simulations utilizing spatially averaged material properties can represent the aging phenomena appreciably, the effect of spatial variability on each early age was also evaluated. Fig.~\ref{marshtcPRV} shows, for the mid-face of the 50.8$\times$50.8 mm (2$\times$2 in) cylinder on WB14/28 days, the spatial distribution of aging related mesoscale LDPM parameters including normal modulus, tensile strength and tensile characteristic length. A direct comparison between simulations with spatially constant (averaged) material properties and spatially variable properties as functions of local aging degree are presented in Fig.~\ref{marsHTC}. For the investigated specimen geometries the differences are quite small with a maximum error in peak loads of 9\%. 

\begin{table}[htbp]
\centering
\begin{tabular}{l c c c c c c}
\multicolumn{7}{l}{Table 6: LDPM Parameters and Material Properties}\\
\hline
Curing\& Age	&	Asym	&HR3d	&HR7d	& HR14d	& HR28d	& WB14/28d \\
		 \hline
Ave. Aging Degree	&1	&0.5582	&0.7061	&0.7762	&0.8103	&0.9541\\
$\pm$Standard Deviation [\%]& & $\pm$0.008 & $\pm$ 0.007 & $\pm$0.072 & $\pm$1.11 & $\pm$0.007 \\
		  \hline

NormalModulus	 [MPa]	&75000	&41865&	52958&	58215&	60773	&71558	\\
DensificationRatio [-]	&	2.5&	2.5	&2.5	&2.5	&2.5	&2.5\\
Alpha [-]		&0.25	&0.25	&0.25	&0.25	&0.25	&0.25\\
TensileStrength [-]	[MPa] 	&13.3&3.4&	5.9&	7.4&	8.1&	11.9\\
CompressiveYieldingStress [MPa]	&	500	&	129&	222&	277&	306&	448\\
ShearStrengthRatio [-]	&5.5&	5.5	&5.5	&5.5&	5.5&	5.5\\
TensileCharacteristicLength [mm] &	10.6&	114&	80&	63&	55&	21\\
SofteningExponent [-]	&	0.28&	0.28&	0.28&	0.28&	0.28&	0.28\\
InitialHardeningModulusRatio [-]		&0.36&	0.36&	0.36	&0.36	&0.36	&0.36\\
TransitionalStrainRatio [-]	&	4&	4&	4	&4	&4&	4	\\
InitialFriction [-]		&0.0335 &0.0335 &0.0335 &0.0335 &0.0335	&0.0335	\\
AsymptoticFriction [-]		&0	&0	&0	&0	&0	&0\\
TransitionalStress [MPa]		&300	&77	&133	&166&	184	&269	\\
VolumetricDeviatoricCoupling [-]		&0	&0&	0	&0	&0	&0	\\
DeviatoricStrainThresholdRatio	 [-]	&1	&1	&1	&1	&1	&1\\
DeviatoricDamageParameter [-]		&5	&5	&5	&5	&5	&5	\\
FractureEnergy [J/m$^2$] & 12.47 & 15.97&	26.28&	29.47&	30.12&	21.21\\
\hline
\end{tabular}
\end{table}

Generally speaking, the age dependent LDPM model is calibrated first by TPB (notched) and cylinder compression test results and later validated by the other types of tests including circular disk Brazilian, cube compression, and unnotched beam bending tests, as well as size effect tests. A thorough discussion of age dependent size effect and the evolution of fracture characteristics can be found in a companion paper \cite{WanUHPCII}. Experimental as well as simulated results of 25.4$\times$25.4$\times$127~mm (1$\times$1$\times$5 in) beam TPB (50\% notched) and unconfined 50.8$\times$50.8 mm (2$\times$2 in) cylinder compression tests can be found in Fig.~\ref{bending} and Fig.~\ref{compression}, respectively. Fig.~\ref{bending}(a) and Fig.~\ref{compression}(a) present specimen setup and typical failure crack pattern from LDPM simulations for TPB and compression tests. In the subfigures, Fig.~\ref{bending}(b-g) and Fig.~\ref{compression}(b-g), the simulation and experimental results on the six investigated early ages are shown. Specimen dimension measurements, ultimate strength, fracture energy, elastic modulus, and number of specimens are reported in the figures as well. The experimental curves of compression tests are modified to eliminate machine compliance by matching the elastic modulus indirectly obtained from beam bending tests, of which CMOD measurements are not affected by the machine deformation or load platen setup. The strength gain and stress-strain curve shapes from simulations match those from experiments very well. The drop in strength from 14 days to 28 days as documented in the experiment section, however, is not captured in the simulations. After 14 days of 100\% RH curing for both curing routines, the specimens are kept at room temperature and room RH of roughly 50\%. The sudden cooling after removing the specimens from the hot water bath likely causes some damage which is further amplified by drying shrinkage cracks. The presented HTC-LDPM early age model, on the other hand, is mainly focused on simulating early age strength gain and mechanical responses of concrete, thus drying shrinkage phenomena as well as the sudden cooling are not included at the moment. 

The mesoscale fracture energy, calculated as $G_t = \ell_{t} \sigma_t^{2} /2E_0$, is also listed in Table 6. 
It is interesting to note that by using the aging formulation presented in this paper, the dependence of the mesoscale fracture energy upon aging degree can be written as follows,
\beq
G_t = \lambda^{2n_a} \frac{l^{\infty}_{t}\sigma_{t}^{\infty 2}}{2E^{\infty}_0}\left(\frac{k_a+1}{\lambda}-k_a\right) 
 \hspace{10mm}  \propto  \hspace{3mm} \lambda^{2n_a}\left(\frac{k_a+1}{\lambda}-k_a\right) 
\eeq
Hence, $G_t$ may or may not reach a maximum for $\lambda < 1$ depending upon the values of $n_a$ and $k_a$, which, as aforementioned, are positive constants.
This means that the fracture energy may exhibit a non-monotonous relationship with the maturity of concrete as documented in the literature for several high performance concretes \cite{Schutter, Ostergaard, Gettu, Kim} and for the UHPC investigated in this study. 
The reason is that according to Hillerborg \cite{Hillerborg85} the fracture energy is dependent on modulus, tensile characteristic length, and tensile strength which may evolve at different rates. 
Since fracture energy is not an independent material property, it must be concluded that the tensile characteristic length is the superior choice for the formulation of aging fracture laws, as demonstrated in this paper. 

Table~7 summarizes the mean values and standard deviations of the experimental as well as simulated nominal splitting tensile strengths for 76.2$\times$25.4~mm (3$\times$1~in) circular disks. Three specimens with different discrete particle configurations are used in the simulations for each age. As discussed earlier, the experimental strengths don't quite follow the expected aging trend, which leads to a certain discrepancy when compared to the strength from aging-functions-based simulations. 

The 25.4~mm (1~in) cube compression tests (Fig.~\ref{cube}) and unnotched 25.4$\times$25.4$\times$127~mm (1$\times$1$\times$5~in) beam TPB (Fig.~\ref{unnotchbeam}) tests are all conducted with 28 days old specimens, which were cured 7 days in the humidity room, 7 days in hot water bath, and then in air for the remaining 14 days. The nominal peak strengths of the unnotched beam bending test simulations lie within the experimental scatter, see Fig.~\ref{unnotchbeam}(b). In both, simulations and experiments, the expected brittle failure mode is observed. Specimen setups, crack patterns, and failure types are presented in the subfigures Fig.~\ref{cube}(b) \& \ref{unnotchbeam}(a). Note that the middle section of the beam, with its length roughly equal to its height, is discretized by lattice discrete particles while the rest of the specimen is a standard finite element mesh with elastic material. This configuration is utilized both in notched and unnotched TPB simulations, see Fig.~\ref{bending} and Fig.~\ref{unnotchbeam}. Since the material in the outer region stays elastic during the entire test, the replacement of LDPM material leads to significant savings in computational time without loss of accuracy. Generally good agreement between simulations and experiments is observed for indirect tensile strength, in particular for the modulus of rupture tests but also, with some limitations, for the Brazilian splitting tests. It is well known that both, Brazilian splitting tests and modulus of rupture tests, are indirect measures of tensile strength. Thus, the splitting tests on 76.2~mm disks, performed after 14 days of HR+WB curing, can be also checked against simulated unnotched bending tests, cured the same way, for which good agreement with the experiments was already documented. In order to exclude bias due to size-effect the check has to be performed on equivalently sized beams, here 6.2$\times$25.4$\times$381~mm (3$\times$1$\times$15~in = height $\times$ width $\times$ length). The predicted modulus of rupture of 14~MPa compares reasonable well with the experimentally obtained splitting tensile strength of 11.4 MPa, further confirming the earlier concluded uncertainties in the experimental splitting tests at later ages. Further deviations between simulation and experiment arise from difficulties in simulating the mechanical boundary conditions of splitting tests, a well known problem. As of the cube compression test, the simulations match the experimental data perfectly, as shown in Fig.~\ref{cube}(a).

\begin{figure} [H]
\centering
\includegraphics[width=6.8in,valign=t]{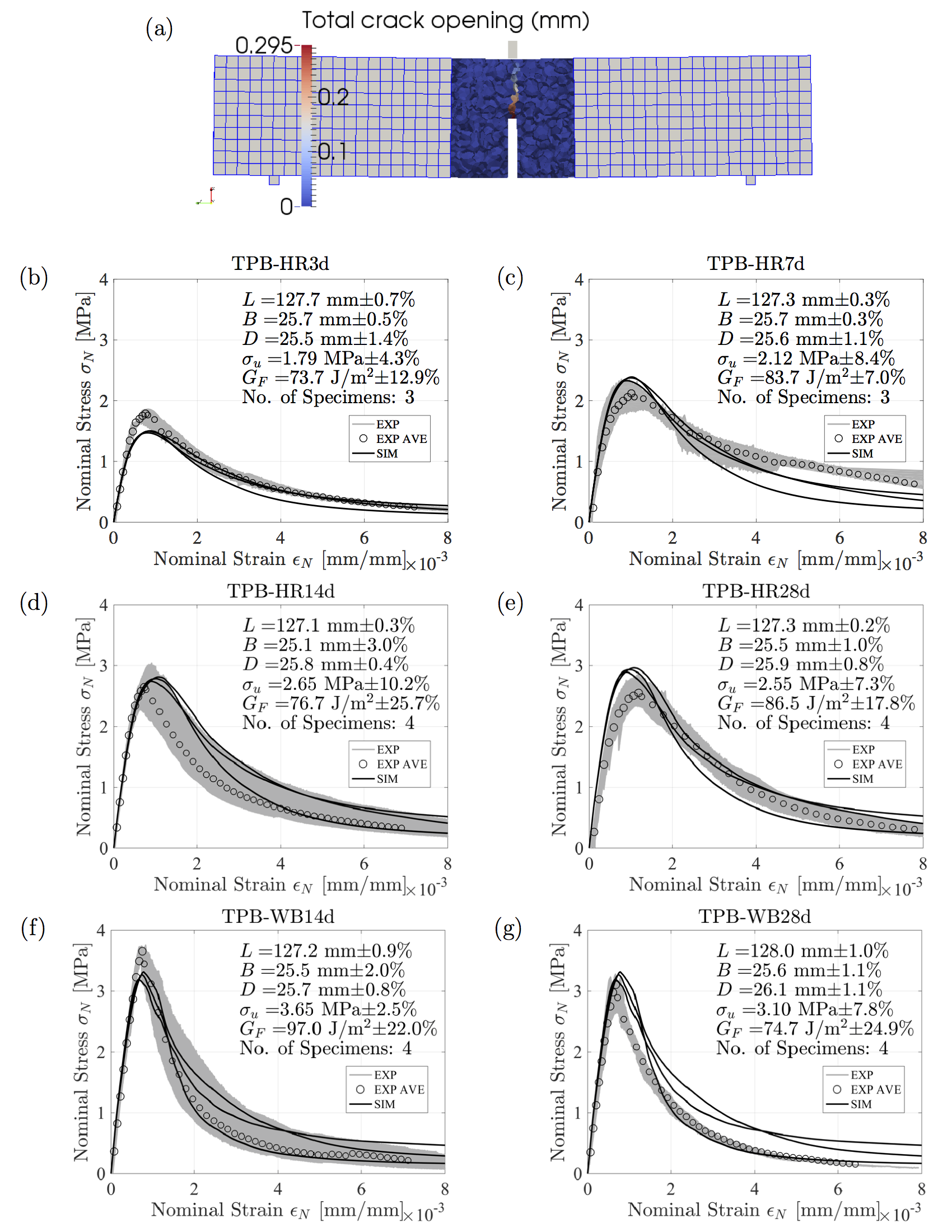}
    \caption{LDPM calibration by experimental results of three-point-bend tests (a) LDPM modeling (b) HR 3 days (c) HR 7 days (d) HR 14 days (e) HR 28 days (f) WB 14 days (g) WB 28 days}
    \label{bending}
\end{figure}

\begin{figure}  [H]
\centering
\includegraphics[width=7in,valign=t]{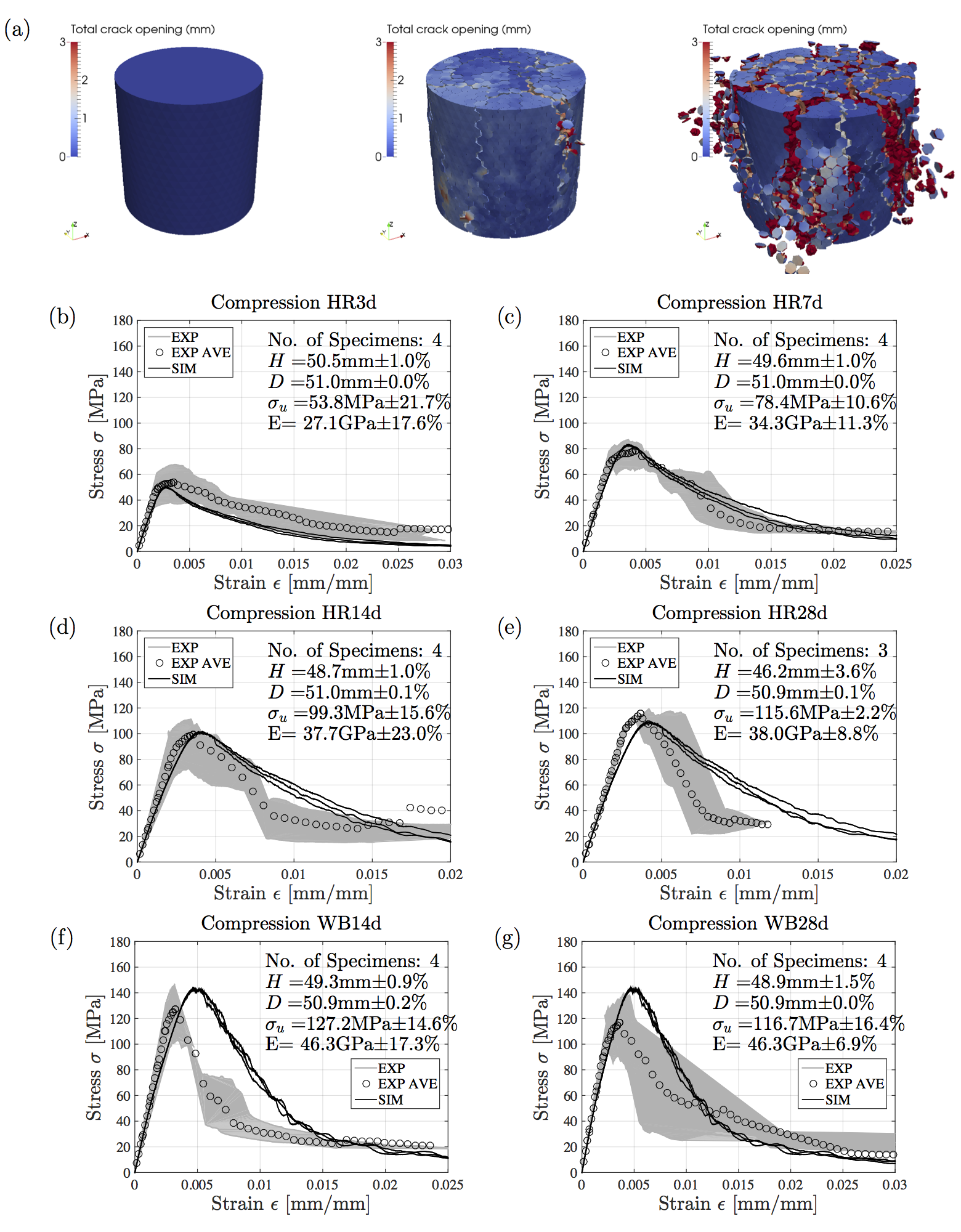}\\
    \caption{LDPM calibration by experimental results of compression tests (a) LDPM modeling (b) HR 3 days (c) HR 7 days (d) HR 14 days (e) HR 28 days (f) WB 14 days (g) WB 28 days}
    \label{compression}
\end{figure}

\begin{figure} [H]
\centering
\includegraphics[width=7in]{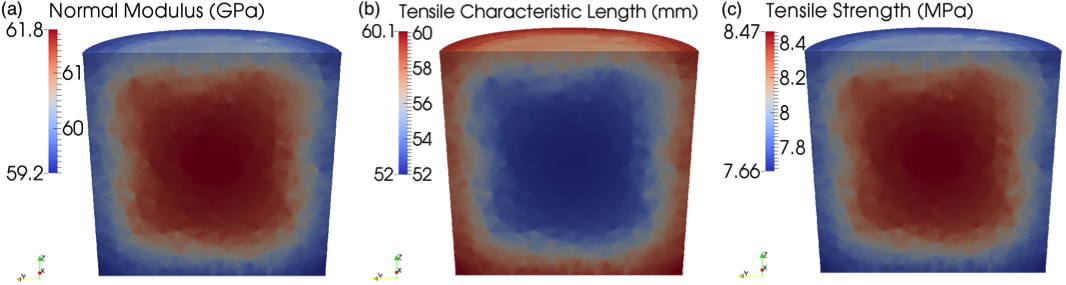}
\caption{Spatial fields of mesoscale LDPM parameters: (a) normal modulus, (b) tensile characteristic length, and (c) tensile strength, on 28 days of age under humidity room (HR) curing \label{marshtcPRV}}   
\label{htc}               
\end{figure}

\begin{figure}  [H]
\centering
   \includegraphics[width=4in]{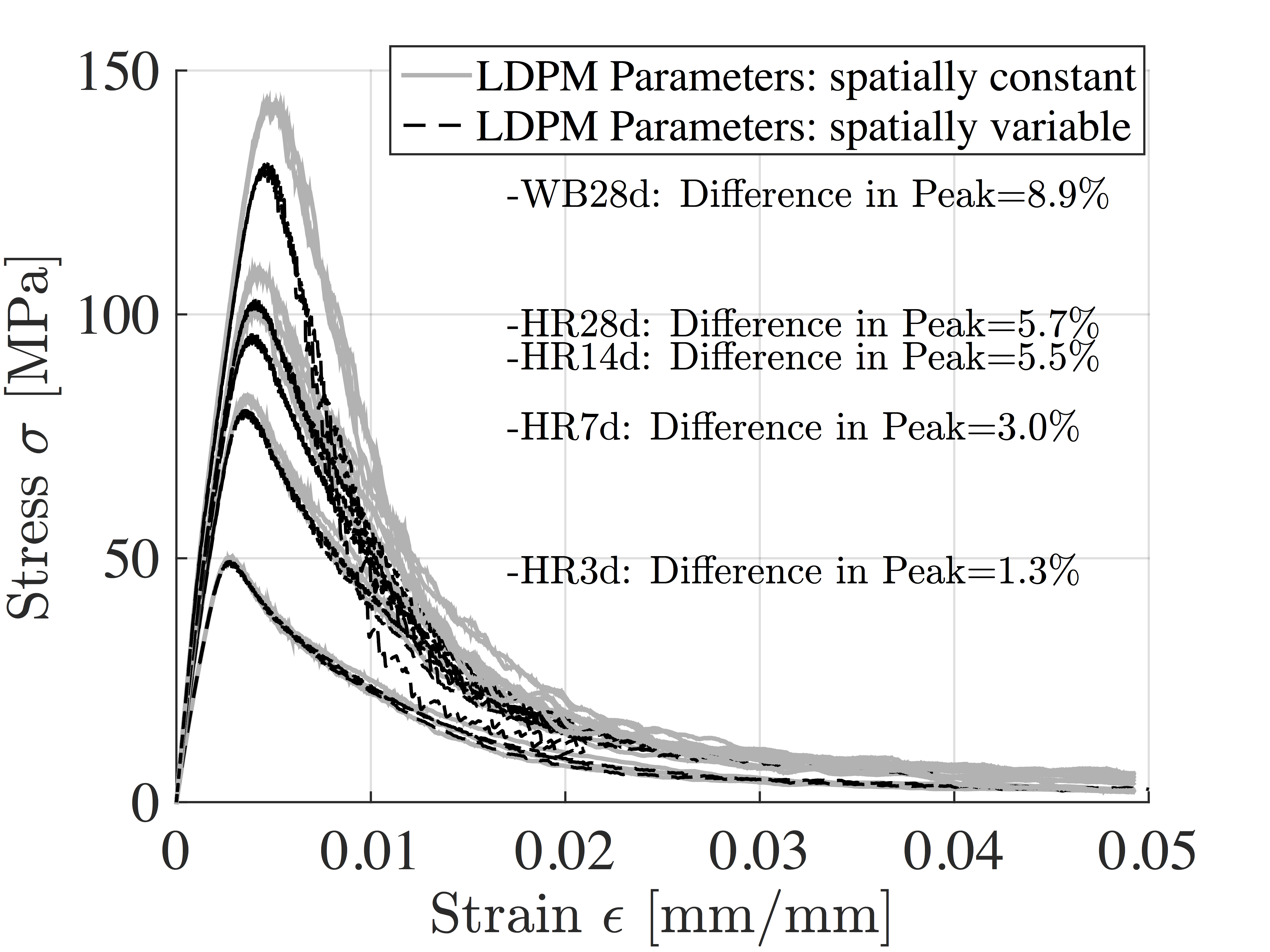}    \hspace{3mm}
    \caption{LDPM simulations with and without spatial variability for compression tests \label{marsHTC}}
\end{figure}

\begin{table}[htbp]
\centering
\begin{tabular}{l c c c c c c}
\multicolumn{7}{c}{Table 7: Brazilian Tensile Strength from Experiments and Simulations}\\
\hline
Age & HR3 & HR7 & HR14 & HR28 & WB14 & WB28\\
\hline
Experiment [MPa] & 4.5 & 5.2 & 5.5 & 4.6 & 11.4 & 9.4\\
 &  $\pm$ 10.4 \% & $\pm$ 9.1 \% & $\pm$ 0 \% & $\pm$ 19.1 \% & $\pm$ 22.7 \% & $\pm$ 12.2 \% \\
 Simulation [MPa] & 3.9 & 6.1 & 6.9 & 7.2 & 7.3 & 7.3\\
&  $\pm$ 3.5 \% &  $\pm$ 3 \% &  $\pm$3.4 \% &  $\pm$3.3 \% &  $\pm$3 \% &  $\pm$3\% \\
 \hline
 \end{tabular}
 \end{table}

\begin{figure} [H]
\includegraphics[height=3in,valign=t]{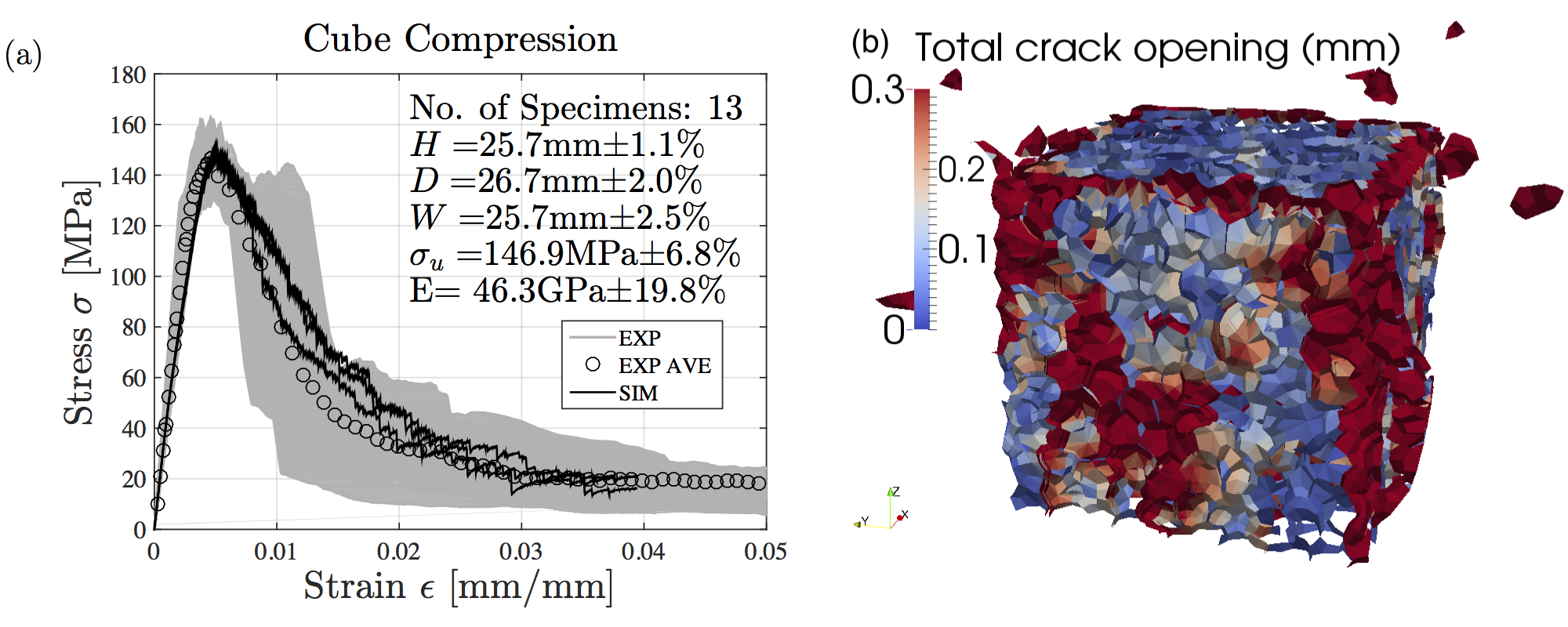}
\caption{LDPM validation by cube compression tests: (a) LDPM simulations and experimental results  (b) LDPM modeling
\label{cube}}
\end{figure}

\begin{figure} [H]
\includegraphics[height=3in,valign=t]{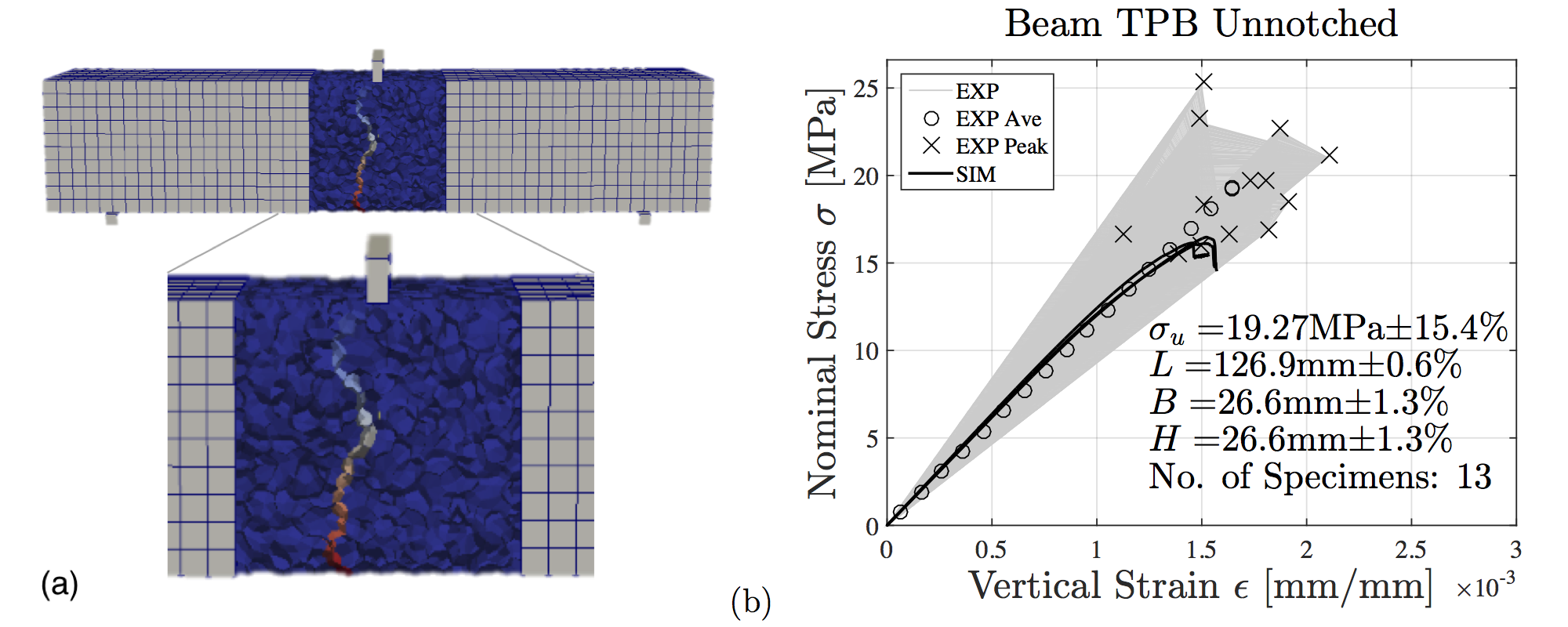}
\caption{LDPM validation by unnotched three-point-bend tests: (a) LDPM modeling setup and crack opening (b) experimental and simulated stress-strain curves 
\label{unnotchbeam}}
\end{figure}

\section{Summary and Conclusions}

In recent years ultra high performance concretes have shown not only a significant rise in popularity but also practical relevance. Yet, a thorough understanding of the evolution of material properties at early age is still lacking in spite of its significance for structural design. In this paper a comprehensive numerical and experimental investigation of the early age behavior of a typical UHPC is presented. The study is based on a large experimental campaign entailing uniaxial compression tests, tensile splitting tests, and three-point-bending tests at different ages and following different curing protocols, complemented by measurements of the evolution of internal humidity in sealed and unsealed samples. 

In order to shed light on the evolution of macroscopic material properties, an early age model is formulated within the framework of mesoscale discrete models. The coupled processes of moisture transport, heat transfer, cement hydration, and silica fume reaction are captured by a hygro-thermo-chemical (HTC) model yielding reaction degrees, which are then used to formulate the aging degree describing the local maturity of the UHPC. The local mesoscale material properties of the constitutive model, the Lattice Discrete Particle Model (LDPM), are obtained through rather simple aging functions, formulated in terms of aging degree.   

Based on the experimental characterization clear trends in the evolution of material properties, namely unconfined compressive strength, tensile splitting strength, flexural strength as well as elastic modulus for the investigated UHPC are noted. These increasing trends are distorted by some of the investigated curing protocols. In particular, the thermal activation in the hot water bath not only accelerates the curing but likely also has adverse affects on the material properties stemming from damage associated with the rapid cooling. Additional distortions in the trends may be attributed to shrinkage damage.

Utilizing the proposed computational early age LDPM framework in furtherance of the comprehensive experimental campaign, the following conclusions can be drawn:
\bi
\item It is possible to obtain post-peak response for UHPC materials, even when they are reaching a mature and, hence, highly brittle state if a closed loop control with a suitable and high accuracy on specimen deformation measurement (e.g. crack opening for fracture or circumferential expansion for cylinder compression) is combined with a sufficiently slow loading rate.
\item In the present investigation thermal activation results in the expected acceleration of the chemical reaction and faster strength gain. However, the often reported improvement in material properties was not observed, on the contrary, possibly due to the negative effects of thermal shock in the aftermath.
\item The maturity of ultra high performance concrete can be predicted accurately if the coupled processes of moisture diffusion, heat transfer, and chemical reactions are captured in a suitable hygro-thermo-chemical framework, taking into account the environmental boundary conditions during curing.
\item The effective aging degree $\lambda$ is a suitable parameter to quantify the maturity of concretes arising from cement hydration and silica-fume reaction, even following different temperature evolutions.
\item The previously postulated linear dependence of Young's modulus on concrete maturity is confirmed. The analysis of the investigated UHPC shows a highly linear dependence of the local mesoscale normal modulus which is related to the macroscopic Young's modulus on aging degree.
\item The evolution of local mesoscale strength-related parameters can be predicted well by monotonously increasing power-law type functions, also formulated in terms of concrete aging degree. The well established empirical square-root relation between Young's modulus and macroscopic (compressive) strength quantities is approximately recovered for the mesoscale properties.  
\item Fracture energy may exhibit a non-monotonous relationship with the maturity of concrete. The reason is that according to Hillerborg the fracture energy is dependent on modulus, tensile characteristic length, and tensile strength, which may evolve at different rates. Since fracture energy is not an independent material property it is not suitable for the formulation of fracture aging laws.
\item Contrary to fracture energy the tensile characteristic length changes monotonously in all investigated cases and shows a linear decreasing dependence on aging degree for the studied UHPC. 
\item The proposed aging framework is capable of describing and predicting the non-monotonous relationship between fracture energy and concrete aging.
\ei
%

Future work includes the extension to and validation by normal strength concretes, and the coupling with creep and shrinkage. This extended LDPM framework will give access to a more substantial inverse analysis of existing datasets and it will allow unrivaled insights into the life-time performance of concrete structures loaded at and exposed to the environment at early age. 

\section*{Acknowledgement}\no

The work of the first and last author was supported under National Science Foundation (NSF) Grant CMMI-1237920 to Northwestern University. The work of the first and second author was also supported by the Austrian Federal Ministry of Economy, Family and Youth, and the National Foundation for Research, Technology and Development. The computational component of this research effort was partially sponsored by the US Army Engineer Research Development Center (ERDC) under Grant W912HZ-12-P-0137. Permission to publish was granted by the director of ERDC geotechnical and structural laboratory. The authors also would like to thank the graduate students Yikai Wang, Allessandro Renzetti and Edgardo Santana for their contributions to the experimental campaign.

\include{reference}
\end{document}

%% file: definitions.tex
\usepackage{amsbsy,amsmath,amsthm,latexsym,amssymb}
\usepackage{graphicx,stmaryrd,sectsty}
\usepackage{setspace}
\usepackage[export]{adjustbox} 
\usepackage[usenames, dvipsnames]{color} 
\definecolor{NU}{RGB}{82,0,99} 

\newcommand{\bc}{\begin{center}}
\newcommand{\ec}{\end{center}}
\newcommand{\bfr}{\begin{flushright}}
\newcommand{\efr}{\end{flushright}}

\newcommand{\no}{\noindent}
\newcommand{\be}{\begin{enumerate}}
\newcommand{\ee}{\end{enumerate}}
\newcommand{\bi}{\begin{itemize}}
\newcommand{\ei}{\end{itemize}}
\newcommand{\bd}{\begin{description}}
\newcommand{\ed}{\end{description}}
\newcommand{\beq}{\begin{equation}}
\newcommand{\eeq}{\end{equation}}
\newcommand{\bea}{\begin{eqnarray}}
\newcommand{\eea}{\end{eqnarray}}

\newcommand{\bfi}{\begin{figure}}
\newcommand{\efi}{\end{figure}}
\newcommand{\bay}{\begin{array}{l}}
\newcommand{\eay}{\end{array}}

\newcommand{\cref}[1]{(\ref{#1})}   

\sectionfont{\normalsize}
\subsectionfont{\small}

\usepackage{color}
\usepackage{tabularx}
\usepackage{float}
\usepackage[section]{placeins}

%% file: segim.tex
\begin{titlepage}
\clearpage\thispagestyle{empty}
\noindent
\hrulefill
\begin{figure}[h!]
\centering
\includegraphics[width=2 in]{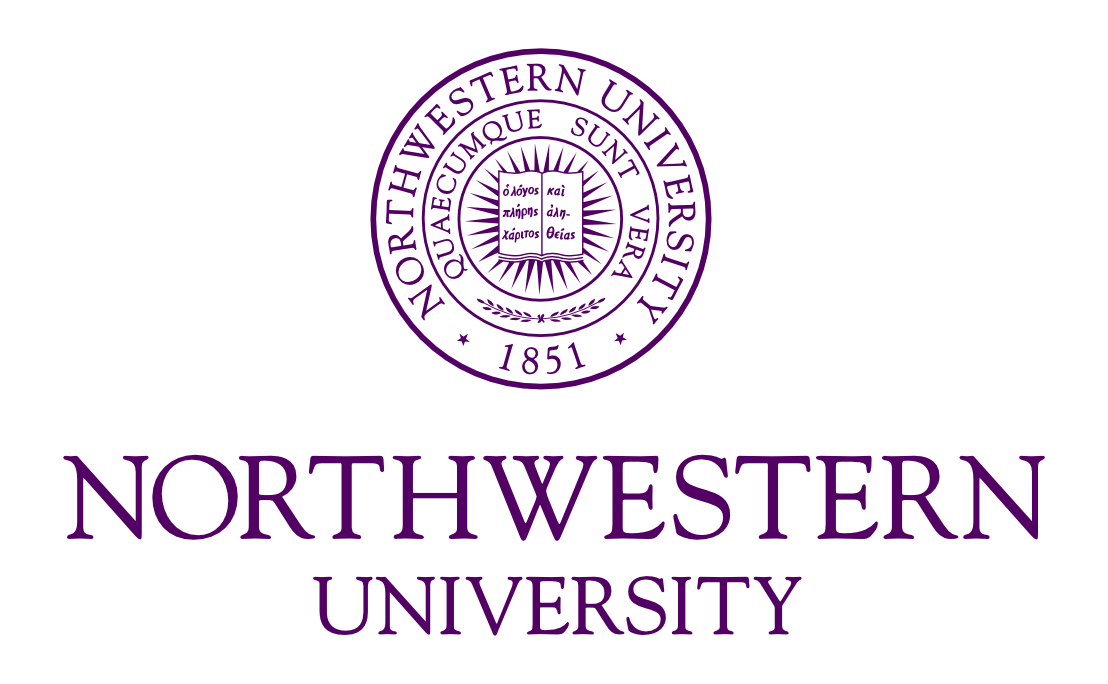}
\end{figure}
\begin{center}
{\color{NU}{
{\bf Center for Sustainable Engineering of Geological and Infrastructure Materials (SEGIM)} \\ [0.1in]
Department of Civil and Environmental Engineering \\ [0.1in]
McCormick School of Engineering and Applied Science \\ [0.1in]
Evanston, Illinois 60208, USA
}
}
\end{center} 
\hrulefill \\ \vskip 2mm
\vskip 0.5in
\begin{center}
{\large {\bf ANALYSIS OF THE BEHAVIOR OF ULTRA HIGH PERFORMANCE CONCRETE AT EARLY AGE}}\\[0.5in]
{\large {\sc Lin Wan, Roman Wendner, Benliang Liang, Gianluca Cusatis}}\\[0.75in]
{\sf \bf SEGIM INTERNAL REPORT No. 15-09/465E}\\[0.75in]
\end{center}
\noindent {\footnotesize {{\em Published in Cement and Concrete Composites \hfill September 2016} }}
\end{titlepage}

\newpage
\clearpage \pagestyle{plain} \setcounter{page}{1}